\documentstyle[12pt]{article}
 
\catcode`\@=11
\long\def\@makefntext#1{
\protect\noindent \hbox to 3.2pt {\hskip-.9pt  
$^{{\ninerm\@thefnmark}}$\hfil}#1\hfill}		

 \def\@makefnmark{\hbox to 0pt{$^{\@thefnmark}$\hss}}  
	
\def\ps@myheadings{\let\@mkboth\@gobbletwo
\def\@oddhead{\hbox{}
\rightmark\hfil\ninerm\thepage}   
\def\@oddfoot{}\def\@evenhead{\ninerm\thepage\hfil
\leftmark\hbox{}}\def\@evenfoot{}
\def\sectionmark##1{}\def\subsectionmark##1{}}
 
\newcounter{sectionc}\newcounter{subsectionc}\newcounter{subsubsectionc}
\renewcommand{\section}[1] {\vspace{0.6cm}\addtocounter{sectionc}{1} 
\setcounter{subsectionc}{0}\setcounter{subsubsectionc}{0}\noindent 
	{\bf\thesectionc. #1}\par\vspace{0.4cm}}
\renewcommand{\subsection}[1] {\vspace{0.6cm}\addtocounter{subsectionc}{1} 
	\setcounter{subsubsectionc}{0}\noindent 
	{\it\thesectionc.\thesubsectionc. #1}\par\vspace{0.4cm}}
\renewcommand{\subsubsection}[1] {\vspace{0.6cm}\addtocounter{subsubsectionc}{1}
	\noindent {\rm\thesectionc.\thesubsectionc.\thesubsubsectionc. 
	#1}\par\vspace{0.4cm}}

\newcounter{appendixc}
\newcounter{subappendixc}[appendixc]
\newcounter{subsubappendixc}[subappendixc]

\renewcommand{\appendix}[1] {\vspace{0.6cm}
        \refstepcounter{appendixc}
        \setcounter{figure}{0}
        \setcounter{table}{0}
        \setcounter{equation}{0}
        \renewcommand{\thefigure}{\Alph{appendixc}.\arabic{figure}}
        \renewcommand{\thetable}{\Alph{appendixc}.\arabic{table}}
        \renewcommand{\theappendixc}{\Alph{appendixc}}
        \renewcommand{\theequation}{\Alph{appendixc}.\arabic{equation}}
        \noindent{\bf Appendix \theappendixc #1}\par\vspace{0.4cm}}

\def\abstracts#1{{
	\centering{\begin{minipage}{30pc}\tenrm\baselineskip=12pt\noindent
	\centerline{\tenrm ABSTRACT}\vspace{0.3cm}
	\parindent=0pt #1
	\end{minipage}}\par}} 
 
\renewenvironment{thebibliography}[1]
	{\begin{list}{\arabic{enumi}.}
	{\usecounter{enumi}\setlength{\parsep}{0pt}
\setlength{\leftmargin 1.25cm}{\rightmargin 0pt}
	 \setlength{\itemsep}{0pt} \settowidth
	{\labelwidth}{#1.}\sloppy}}{\end{list}}
 
\newcounter{itemlistc}
\newcounter{romanlistc}
\newcounter{alphlistc}
\newcounter{arabiclistc}

\newcommand{\fcaption}[1]{
        \refstepcounter{figure}
        \setbox\@tempboxa = \hbox{\tenrm Fig.~\thefigure. #1}
        \ifdim \wd\@tempboxa > 6in
           {\begin{center}
        \parbox{6in}{\tenrm\baselineskip=12pt Fig.~\thefigure. #1}
            \end{center}}
        \else
             {\begin{center}
             {\tenrm Fig.~\thefigure. #1}
              \end{center}}
        \fi}
 
\newcommand{\tcaption}[1]{
        \refstepcounter{table}
        \setbox\@tempboxa = \hbox{\tenrm Table~\thetable. #1}
        \ifdim \wd\@tempboxa > 6in
           {\begin{center}
        \parbox{6in}{\tenrm\baselineskip=12pt Table~\thetable. #1}
            \end{center}}
        \else
             {\begin{center}
             {\tenrm Table~\thetable. #1}
              \end{center}}
        \fi}
 
\def\@citex[#1]#2{\if@filesw\immediate\write\@auxout
	{\string\citation{#2}}\fi
\def\@citea{}\@cite{\@for\@citeb:=#2\do
	{\@citea\def\@citea{,}\@ifundefined
	{b@\@citeb}{{\bf ?}\@warning
	{Citation `\@citeb' on page \thepage \space undefined}}
	{\csname b@\@citeb\endcsname}}}{#1}}
 
\newif\if@cghi
\def\cite{\@cghitrue\@ifnextchar [{\@tempswatrue
	\@citex}{\@tempswafalse\@citex[]}}
\def\citelow{\@cghifalse\@ifnextchar [{\@tempswatrue
	\@citex}{\@tempswafalse\@citex[]}}
\def\@cite#1#2{{$\null^{#1}$\if@tempswa\typeout
	{IJCGA warning: optional citation argument 
	ignored: `#2'} \fi}}

\def\fnt#1#2{\footnotetext{\kern-.3em
	{$^{\mbox{\sevenrm #1}}$}{#2}}}
 
 1
 1
 1

\font\tenbf=cmbx10
\font\tenrm=cmr10
\font\tenit=cmti10

\font\ninerm=cmr9

\begin{document}



 \newcommand{\be}[1]{\begin{equation}\label{#1}}
 \newcommand{\ee}{\end{equation}}
 \newcommand{\beqn}[1]{\begin{eqnarray}\label{#1}}
 \newcommand{\eeqn}{\end{eqnarray}}
\newcommand{\bd}{\begin{displaymath}}
\newcommand{\ed}{\end{displaymath}}
\newcommand{\mat}[4]{\left(\begin{array}{cc}{#1}&{#2}\\{#3}&{#4}\end{array}
\right)}
\newcommand{\matr}[9]{\left(\begin{array}{ccc}{#1}&{#2}&{#3}\\
{#4}&{#5}&{#6}\\{#7}&{#8}&{#9}\end{array}\right)}
\def\la{\lambda}
\def\al{\alpha}
\def\Ga{\Gamma}
\def\ga{\gamma}
 \newcommand{\eps}{\varepsilon}
\newcommand{\ov}{\overline}
\renewcommand{\to}{\rightarrow}
\def\mcirc{{\stackrel{o}{m}}}
%
%
\makeatletter
\newcounter{alphaequation}[equation]
\def\thealphaequation{\theequation\alph{alphaequation}}
%
\def\eqnsystem#1{
\def\@eqnnum{{\rm (\thealphaequation)}}
\def\@@eqncr{\let\@tempa\relax
\ifcase\@eqcnt \def\@tempa{& & &}
\or \def\@tempa{& &}\or \def\@tempa{&}\fi\@tempa
\if@eqnsw\@eqnnum\refstepcounter{alphaequation}\fi
\global\@eqnswtrue\global\@eqcnt=0\cr}
\refstepcounter{equation}
\let\@currentlabel\theequation
\def\@tempb{#1}
\ifx\@tempb\empty\else\label{#1}\fi
\refstepcounter{alphaequation}
\let\@currentlabel\thealphaequation
\global\@eqnswtrue\global\@eqcnt=0
\tabskip\@centering\let\\=\@eqncr
$$\halign to \displaywidth\bgroup
  \@eqnsel\hskip\@centering
  $\displaystyle\tabskip\z@{##}$&\global\@eqcnt\@ne
  \hskip2\arraycolsep\hfil${##}$\hfil&
  \global\@eqcnt\tw@\hskip2\arraycolsep
  $\displaystyle\tabskip\z@{##}$\hfil
  \tabskip\@centering&\llap{##}\tabskip\z@\cr}

\def\endeqnsystem{\@@eqncr\egroup$$\global\@ignoretrue}
\makeatother


\begin{flushright}
hep-ph/9602325 ~~~~~ INFN-FE 21/95 \\ 
December 1995 \\
\end{flushright}
\vspace{10mm}

\centerline{\tenbf FERMION MASSES AND MIXING IN SUSY GUT 
\footnote{Based on lectures delivered at the ICTP Summer 
School {\em on High Energy Physics and Cosmology}, Trieste, Italy, 
3-28 July 1995 (to appear on Proceedings), 
and XIX Int. Conference on {\em Particle Physics and Astrophysics 
in the Standard Model and Beyond}, Bystra, Poland, 19-26 September 
1995 } }

\vspace{0.8cm}
\centerline{\tenrm ZURAB BEREZHIANI }
\baselineskip=13pt
\centerline{\tenit INFN Sezione di Ferrara, 44100 Ferrara, Italy,}
\baselineskip=12pt
\centerline{\tenit Institute of Physics, Georgian Academy of Sciences, 
380077 Tbilisi, Georgia} 
\vspace{0.9cm}

\abstracts{
The problem of fermion masses and mixings is discussed in the context 
of supersymmetric grand unification theories. 
Some predictive frameworks
based on the $SU(5)$, $SO(10)$ and $SU(6)$ models are reviewed. }
 
\vfil
\rm\baselineskip=14pt

\section{Introduction}
\vspace{-0.7cm}
\subsection{Family Problems }
\vspace{-0.35cm}

The problem of fermion flavours (or families) is one of the key 
problems in modern particle physics. It has different aspects, 
questioning origin of the family replication (why three families?),  
quark and lepton mass spectrum and mixing pattern, CP violation in 
weak interactions,  CP conservation in strong interactions,  
suppression of the flavour changing neutral currents (FCNC), 
pattern of neutrino masses and oscillations, etc. 

The Standard Model\cite{SM} (SM) can be considered as a minimal 
theory of flavour. Being an internally consistent renormalizable 
gauge theory, it has been extremely successfull in describing 
various experimental data accumulated over the past several years. 
It is likely that the SM is a literally correct theory 
at presently available energies. 
It accomodates all observed quarks and leptons in a consistent way. 
Three families sharing the same quantum numbers under 
the $SU(3)\times SU(2)\times U(1)$ gauge symmetry 
are introduced as an anomaly free set of chiral 
left-handed (LH) fermions $q_i=(u_i,d_i)$, $u^c_i$, $d^c_i$; 
$l_i=(\nu_i,e_i)$, $e^c_i$,  where $i=1,2,3$ is a family index 
($q,l$ are the weak isodoublets, and the isosinglet states 
 $u^c,d^c,e^c$ are the 
C-conjugates to the right-handed (RH) components $u_R,d_R,e_R$).  
A remarkable feature of the SM is that the fermion and 
the gauge boson $W^{\pm},Z$ masses have a common origin, 
namely the Higgs mechanism. In fact, fermions would remain 
massless as far as gauge symmetry is unbroken. They get masses 
through the Yukawa couplings to the Higgs doublet $\phi$: 
\be{YSM} 
{\cal L}_{\rm Yuk} = \la^u_{ij} q_iCu^c_j \tilde{\phi}\,+\,
\la^d_{ij} q_iCd^c_j \phi \, + \,\la^e_{ij}l_iCe^c_j \phi   
~~~~~~~~(\tilde{\phi}=i\tau_2 \phi^{\ast})
\ee
So, the fermion masses are related to the weak scale 
$\langle \phi \rangle=v=174$ GeV. However, the Yukawa constants remain 
arbitrary: $\hat{\la}^{u,d,e}$ are general complex $3\times 3$ matrices.

The SM contains no renormalizable couplings that could generate the 
neutrino masses. As far as renormalizable interactions are concerned,  
the lepton and baryon number conservations arise as accidental 
symmetries of the theory. The lowest order couplings relevant for the 
neutrino masses are the $d=5$ ones\cite{Weinb}
\be{Ynu-SM}
{\cal L}_{\nu} =
\frac{\la^{\nu}_{ij}}{M} (l_i \tilde{\phi})C(l_j \tilde{\phi})\,, 
~~~~~~~~~~~~~~ \la^{\nu}_{ij}=\la^{\nu}_{ji}
\ee 
where $M\gg v$ is some regulator scale. In the seesaw 
picture\cite{seesaw}
these can be induced by exchange of the heavy RH neutrinos. 
In this case $M$ is related to the mass scale of the latter. 
Alternatively, if the SM is valid all the way up to planckian energies  
(i.e. there are no RH neutrinos with mass $\leq M_{Pl}$), then 
operators (\ref{Ynu-SM}) with $M\sim M_{Pl}$ could effectively emerge 
due to the non-perturbative quantum gravitational effects\cite{BEG}. 
In this view, the value $\mcirc = v^2/M_{Pl}=3\cdot 10^{-6}$ eV 
can be regarded as a natural unit of the neutrino masses in the SM.

The coupling constant matrices and correspondingly the fermion 
mass matrices $\hat{m}^f=\hat{\la}^fv$ ($f=u,d,e$) and   
$\hat{m}^\nu=\hat{\la}^\nu (v^2/M)$ can be brought to the 
diagonal form by the unitary transformations: 
\be{unit}
V_{f}\hat{m}^fV'_{f}=\hat{m}^f_{diag}\,, ~~~~~~~~~
V_{\nu}\hat{m}^\nu V^T_\nu =\hat{m}^\nu_{diag}
\ee
where 
\beqn{diag}
&&
\hat{m}^u_{diag}=\mbox{diag}(m_u,\,m_c,\,m_t)
=v\cdot \mbox{diag}(\la_u,\,\la_c,\,\la_t) \nonumber \\
&&
\hat{m}^d_{diag}=\mbox{diag}(m_d,\,m_s,\,m_b) 
=v\cdot \mbox{diag}(\la_d,\,\la_s,\,\la_b) \nonumber \\
&&
\hat{m}^e_{diag}=\mbox{diag}(m_e,\,m_\mu,\,m_\tau)
=v\cdot \mbox{diag}(\la_e,\,\la_\mu,\,\la_\tau)  \nonumber \\
&&
\hat{m}^\nu_{diag}=\mbox{diag}(m_1,\,m_2,\,m_3)
=\frac{v^2}{M}\cdot \mbox{diag}(\la_1,\,\la_2,\,\la_3) 
\eeqn
Hence, quarks are mixed in the charged current interactions: 
\be{mix}
{\cal L}_W=\frac{g}{\sqrt{2}}\overline{(u_1,u_2,u_3)}_L\gamma^{\mu}W^+_{\mu}
\left( \begin{array}{c} d_1\\d_2\\d_3 \end{array} \right)_L=
\sqrt{2}g\,\overline{(u,c,t)}\,\gamma^{\mu}(1+\gamma^5)W^+_{\mu}V_{\rm CKM}
\left( \begin{array}{c} d\\s\\b \end{array} \right) 
\ee
where $V=V^+_{u}V_{d}$ is the Cabibbo-Kobayashi-Maskawa (CKM) 
matrix. In a convenient parametrization it has a form 
\be{CKM}
V_{\rm CKM} = \matr{c_{12}c_{13}}{s_{12}c_{13}}{s_{13}e^{-i\delta}}
{-s_{12}c_{23}-c_{12}s_{23}s_{13}e^{i\delta}}
{c_{12}c_{23}-s_{12}s_{23}s_{13}e^{i\delta}} {s_{23}c_{13}} 
{s_{12}s_{23}-c_{12}c_{23}s_{13}e^{i\delta}}
{-c_{12}s_{23}-s_{12}c_{23}s_{13}e^{i\delta}} {c_{23}c_{13}} 
\ee
where $c_{ij}$ and $s_{ij}$ respectively stand for the `$\cos$' and 
`$\sin$' of the mixing angles $\theta_{12}$, $\theta_{23}$
and $\theta_{13}$, and $\delta$ is a CP-violating phase. 
In the case of massive neutrinos, a similar mixing matrix 
$V_{\rm Lept}=V^+_\nu V_e$ emerges also in the lepton sector.

The mass spectrum of the quarks and charged leptons is spread over 
five orders of magnitude, from MeVs to 100 GeVs:\cite{Datagroup}: 
\beqn{masses}
&&
m_t=170\pm 12\,{\rm GeV}, ~~~
m_c=1.3\pm 0.1\,{\rm GeV}, ~~~~
m_u=2-8\,{\rm MeV}    \nonumber \\
&&
m_b=4.3\pm 0.2\,{\rm GeV}, ~~~
m_s=100-300\,{\rm MeV}, ~~~
m_d=5-15\,{\rm MeV}  \nonumber \\
&&
m_\tau=1.784\,{\rm GeV}, ~~~~~~~
m_\mu=105.6\,{\rm MeV}, ~~~~~~~~
m_e=0.511\,{\rm MeV}
\eeqn
Following the tradition\cite{GL}, for the heavy quarks $t,b,c$,
we refer to their running masses respectively at $\mu=m_{t,b,c}$,  
and for the light quarks $u,d,s$ -- at $\mu=1$ GeV . 
For the top quark `pole' mass $M_t= m_t[1+(4/3\pi)\alpha_3(m_t)]$  
the recent results of the CDF and D0 groups\cite{CDF} imply respectively 
$M_t=176 \pm 8\pm 10$ GeV and $M_t=199 \pm 20\pm 22$ GeV, 
with the average $M_t=180 \pm 12$ GeV. This is in a good agreement 
with the present precision data on the SM. 

The light quark masses are the less known quantities in (\ref{masses}), 
however their ratios are known with the better accuracy\cite{Leut}: 
\be{m-ratios}
\frac{m_u}{m_d}=0.25 - 0.70, ~~~~~ \frac{m_s}{m_d}=17-25; 
~~~~~~
\left(\frac{m_u}{m_d}\right)^2+ 
\frac{1}{Q^2}\left(\frac{m_s}{m_d}\right)^2 = 1 ~~~~(Q=23\pm 2) 
\ee

The weak transitions dominantly occur inside the families, and are 
suppressed between different families. 
For the CKM mixing angles we have\cite{Datagroup} 
\beqn{angles} 
&& |V_{us}|=s_{12}=0.222\pm 0.002 ~~~~~ \nonumber \\ 
&& |V_{cb}|=s_{23}=0.040\pm 0.005 ~~~~~ \nonumber \\ 
&& |V_{ub}|=s_{13}=(0.08\pm 0.02)\cdot s_{23} 
\eeqn  

Direct measurements show no evidence for any of the neutrinos to be 
massive, providing only the upper bounds\cite{Datagroup}:  
\be{nu-mass} 
m_{\nu_\tau}< 31 ~{\rm MeV}, ~~~~ 
m_{\nu_\mu}< 270 ~{\rm keV}, ~~~~ 
m_{\nu_e}< 7.0 ~{\rm eV} ~~~~ 
[2\beta_{0\nu}: ~~ m^\nu_{ee}<0.7 ~{\rm eV}] 
\ee
On the other hand, there have been indirect ``positive'' signals for 
neutrino masses and mixing accumulated during the past years. 
The most serious hint among these is related to the 
{\em solar neutrino problem} (SNP). 
A solar neutrino deficit indicated by the current experimental data  
cannot be explained by nuclear/astrophysical reasons\cite{SNP}. 
This points that the SNP is rather due to the neutrino properties, 
the most natural and plausible solution being the solar $\nu_e$ 
oscillation into another neutrino $\nu_x$ ($\nu_x=\nu_\mu$ or $\nu_\tau$).  
It can explain the experimental data in two following regimes. 

{\em Just-so}\cite{just-so}: long wavelength oscillation from 
Sun to Earth, with parameters in the range 
\be{JS}
\delta m^2_{ex}\sim 10^{-10}~{\rm eV}^2, ~~~~~ 
\sin^2 2\theta_{ex}\sim 1 
\ee 
i.e. neutrino masses $\sim \mcirc$ and almost maximal mixing. 
In the context of the operators (\ref{Ynu-SM}) this implies 
that $M\sim M_{Pl}$ and all $\la^\nu_{ij}\sim 1$, as it 
could emerge from the Planck scale physics\cite{BEG}.

{\em MSW}\cite{MSW}: resonant oscillation inside the solar medium, 
with the parameter range
\be{MSW}
\delta m^2_{ex}\sim 10^{-5}~{\rm eV}^2, ~~~~~ 
\sin^2 2\theta_{ex}\sim 10^{-2}  
\ee 
This case favours smaller scale $M\sim 10^{12-16}$ GeV, and 
the ``hierarchial'' form of the matrix $\la^\nu_{ij}$ alligned to 
the charged leptons Yukawa matrix $\la^{e}_{ij}$. 

Other hints, as are the atmospheric neutrino 
deficit\cite{ANP}, LSND anomaly\cite{LSND} or the "after COBE" evidence 
for some hot fraction of the cosmological dark matter\cite{HDM}, 
point to heavier ($m_\nu\sim 0.1-10$ eV) and substantially 
mixed neutrinos. 

Summarizing, one can conclude that the SM accomodates the fermion 
sector in a consistent way. There is only one dimensional parameter, 
$v=174$ GeV, which determines the mass scale of the charged fermions.  
It is of key importance that the SM exhibits the {\em natural} 
suppression of the flavour changing neutral currents (FCNC), both in 
the gauge boson and Higgs exchanges\cite{FCNC}. 
Since the Yukawa constants in (\ref{YSM}) are generally 
complex, the observed CP-violating phenomena can be explained 
by the CKM mechanism with sufficiently large CP-phase 
($\delta\sim 1$). However, at the same time this creates the strong 
CP problem\cite{StrCP}: 
the overall phase of the complex Yukawa matrices would effectively 
contribute to the $\Theta$-term in QCD and thus 
induce the P and CP violation in strong interactions. On the other 
hand, absence of the dipole electric moment of neutron puts 
a strong bound $\Theta < 10^{-9}$.

\subsection{Fermion masses beyond the Standard Model }
\vspace{-0.35cm}

As noted above, the fermion mass and mixing problem can be 
phrased as a problem of the Yukawa coupling matrices $\hat{\la}^f$ 
which remain arbitrary in the SM: there is no 
explanation, what is the origin of a strong hierarchy between their 
eigenvalues, why $\hat{\la}^u$ and $\hat{\la}^d$ are 
alligned so that the CKM mixing angles are small, what is the origin of 
the complex structure needed for the CP-violation in weak interactions, 
why the $\Theta$-term is vanishingly small in spite of the 
complex Yukawas, etc. 

It is attractive to think that at some intermediate scale $M_F$ 
between the electroweak and Planck scales there exists a more 
fundamental theory which could allow to calculate the Yukawa  
couplings, or at least somehow constrain them. 

In order to analyze predictions of such a 
theory, one has to compare the quark and lepton running masses 
at the scale $\mu\sim M_F$. The latter are related to the `physical' 
masses (\ref{masses}) through the renormalization group (RG) equations. 
One has to remember, however, that these equations contain 
a principally unknown parameter:  
\be{X}
{\cal X}(M_F)= \mbox{Spectrum of particles below the scale $M_F$}  
\ee
The minimal assumption is that ${\cal X}(M_F)={\cal X}_{\rm SM}$, 
i.e. below the scale $M_F$ the theory reduces to the minimal SM 
literally in all its sectors (gauge, fermion, Higgs)  
and there are no extra particles besides the known ones.  
(In the context of the MSSM there will be also 
their superpartners with masses $m_S\sim 1$ TeV). 
In this case the RG running is essentially defined by the gauge 
coupling constants $g_3,g_2,g_1$ of the $SU(3)\times SU(2)\times U(1)$ 
symmetry and large top Yukawa constant ($\la_t\sim 1$). 
The effects of large $\la_t$ imply the existence of the infrared-fixed 
point\cite{PR} and also influence the RG running of other 
Yukawa constants and mixing angles\cite{OP}. 


Looking on the fermion mass spectrum at the scale $\mu\sim M_F\gg M_W$, 
we observe that it is divided into 
following groups (in units of the weak scale $v=174$ GeV): 
\be{pattern}
m_t\sim v, ~~~~ m_{c,b,\tau}\sim 10^{-2} v, ~~~~ 
m_{s,\mu}\sim 10^{-3} v,  ~~~~ m_{u,d,e}\sim 10^{-5} v  
\ee 
One can also observe that the {\em vertical} mass splitting is small within 
the first family of quarks and is quickly growing with the family number:
\be{vert}
\frac{m_u}{m_d}\sim 1\,,~~~~
\frac{m_c}{m_s}\sim 10\,,~~~~
\frac{m_t}{m_b}\sim 10^2 
\ee
whereas the mass splitting between the charged leptons 
and down quarks remains considerably smaller: 
\be{vert-lept}
\frac{m_e}{m_d}\sim 0.3\,,~~~~
\frac{m_\mu}{m_s}\sim 3\,,~~~~
\frac{m_\tau}{m_b}\sim 1  
\ee
so that at large $\mu$ the third family is almost unsplit, 
$m_b\simeq m_\tau$, whereas the first two families are split 
but $m_d m_s\sim m_e m_\mu$: 

{\em Horizontal} hierarchy of quark 
masses exhibites the approximate scaling low 
\be{qh} 
m_t:m_c:m_u\sim 1:\eps_u:\eps_u^2\,,~~~~~~~ 
m_b:m_s:m_d\sim 1:\eps_d:\eps_d^2\,
\ee 
where $\eps_u^{-1}=200-300$ and $\eps_d^{-1}=20-30$. The charged 
leptons masses have a mixed behaviour: 
\be{lh}
m_\tau :m_\mu : m_e\sim 1:\eps_e:\eps_e\eps'_e 
\ee 
where $\eps_e\sim \eps_d$ and $\eps'_e\sim \eps_u$. 
One can also exploit experimental information on the quark mixing.  
The CKM angles exhibit the following hierarvhy: 
\be{mix-eps} 
s_{12}\sim \eps_d^{1/2}\,,~~~~ s_{23}\sim \eps_d\,,~~~~
s_{13}\sim \eps_d^{2}\,,
\ee
which points to the correlations between the quark mass spectrum 
and mixing pattern. Moreover, there are intriguing relations 
between masses and mixing angles,  such as the 
well-known formula for the Cabibbo angle $\,s_{12}=\sqrt{m_d/m_s}\,$.  

It is tempting to think that correlation between the mixing angles and 
fermion masses is intrinsically connected to the peculiarities of the 
hypothetical flavour physics at the scale $\mu\sim M_F$, and 
and the former actually are the functions of the latter. 
It is also suggestive that the CKM angles have the 
following "analytic" properties\cite{decoupling}:  

\underline{\em  Decoupling}. Mixings of the first 
family with others ($s_{12}$, $s_{13}$) vanish in the limit 
$m_u,m_d\rightarrow 0$. At the next step, 
when $m_c,m_s\rightarrow 0$, $s_{23}$ also vanishes. 

\underline{\em Scaling}. All mixing angles $s_{12},s_{13},s_{23}$ 
vanish in the limit when masses of the up and down quarks are 
proportional to each other: $m_u:m_c:m_t=m_d:m_s:m_b$.

\newpage

\section{ Fermion Masses in SUSY GUT: the SU(5) Lessons } 
\vspace{-0.7cm}
\subsection{From Love Story to Family Problems}
\vspace{-0.35cm}

Nowadays the most promising ideas beyond the SM are related to 
the concepts of supersymmetry (SUSY)\cite{SUSY}  and 
grand unification theories (GUT)\cite{su5,Pati,so10}.   
Their relations can be expressed by a simple formula 
\be{Love}
\mbox{SUSY} + \mbox{GUT} 
=~~_{/}\!\!\! \heartsuit\!\!^{^\nearrow}
\ee
Softly broken (at the scale $m_S\sim 1$ TeV) supersymmetry  
is the only plausible idea that can support the GUT 
against the gauge hierarchy problem\cite{Maiani}. 
The {\em present} data on $\al_3(M_Z)$ and 
$\sin^2\theta_W(M_Z)$ are in a remarkable agreement with 
the {\em elder} prediction\cite{DRW} of the SUSY $SU(5)$  
while exclude the non-supersymmetric $SU(5)$.\cite{Amaldi}   
On the other hand, the minimal supersymmetric standard 
model (MSSM) without GUT is also not in best shape: 
unification at the string scale gives too small $\sin^2\theta_W(M_Z)$. 
In the MSSM the running gauge constants $g_3$, $g_2$ and $g_1$ 
given at $\mu=M_Z$ withing their experimental error bars, in their 
evolution to higher energies join at the scale 
$M_X\simeq 10^{16}$ GeV.\cite{Amaldi}  
Hence, at this scale the $SU(3)\times SU(2)\times U(1)$ symmetry 
can be consistently embedded into $SU(5)$, 
which at larger scales can be further extended to larger groups. 
At this point the elegant GUT and beautiful SUSY,  
already long time attracted to each other, finally successfully meet.

All these suggest a following paradigm: a basic (string?) ``Theory 
of Everything'' below the Planck scale $M_P$ reduces\footnote{
Without knowing exactly to what is the basic scale of the theory, 
in the following under the Planck scale we imply a broad range 
$M_P\sim 10^{17-19}$ GeV, unless it is specified. In particular,  
this can be the Planck mass  $M_{Pl}\simeq 10^{19}$ GeV itself, 
reduced Planck mass $2\cdot 10^{18}$ GeV, or a 
string scale $\sim 3\cdot 10^{17}$ GeV. }
~to a SUSY GUT containing the $SU(5)$
subgroup, which then at $M_X\simeq 10^{16}$ GeV
breaks down to the $SU(3)\times SU(2)\times U(1)$. 
Below the scale $M_X$ starts  {\em Great Desert}, with no extra 
particles  besides the known ones and their superpartners. 
In other words, ${\cal X}(M_X)={\cal X}_{\rm MSSM}$, 
where ${\cal X}_{\rm MSSM}$ denotes the MSSM particle content 
containing the chiral superfields of quarks and leptons 
$q_i$, $l_i$, $u^c_i$, $d^c_i$, $e^c_i$ ($i=1,2,3$), two 
Higgs doublets $\phi_{1,2}$, and the gauge superfields of 
$SU(3)\times SU(2)\times U(1)$. In fact, some extra complete degenerate 
supemultiplets could populate intermediate scales without spoiling  
unification of the gauge constants. However, this would 
affect the RG factors in the Yukawa constant running 
(see below, eq. (\ref{RG})). 

In the MSSM the fermion masses emerge from the superpotential terms 
\be{Y-MSSM} 
{\cal W}_{\rm Yuk}= \la^u_{ij} q_i u^c_j \phi_2\,+\,
\la^d_{ij} q_i d^c_j \phi_1 \,+ \,\la^e_{ij} l_i e^c_j \phi_1,    
~~~~~~~ 
{\cal W}_\nu= \frac{\la^{\nu}_{ij}}{M_P} (l_i \phi_2)(l_j \phi_2)
\ee
which are strightforward extension of the SM couplings (\ref{YSM}) 
and (\ref{Ynu-SM}).   
The Yukawa matrices $\la^{u,d,e,\nu}$ remain arbitrary in the MSSM,  
while  presence of two Higgses $\phi_1$ and  $\phi_2$ with VEVs 
$v_1=v\cos\beta$ and $v_2=v\sin\beta$ ($v=174$ GeV)   
involves also an additional parameter $\tan\beta=v_2/v_1$. 

Applied to the flavour problem, grand unification can play 
an important role in understanding the fermion mass spectrum. 
It can allow to calculate the Yukawa constants at the scale 
$M_X$, or at least somehow constrain them.  
In order to confront these predictions  
to the observable mass pattern (\ref{masses}), one has to 
account for the Yukawa constants RG running down from the scale 
$M_X\sim 10^{16}$ GeV. By assuming that 
${\cal X}(M_X)={\cal X}_{\rm MSSM}$ and considering 
moderate values of $\tan\beta$, one obtains\cite{Barger}:
\beqn{RG}
&&
m_t= \lambda_t A_u y^6 v\sin\beta, ~~~~
m_c= \lambda_c A_u \eta_c y^3 v\sin\beta, ~~~~ 
m_u= \lambda_u A_u \eta y^3 v\sin\beta    \nonumber \\
&&
m_b= \lambda_b A_d \eta_b y v\cos\beta , ~~~~
m_s= \lambda_s A_d \eta v\cos\beta, ~~~~
m_d= \lambda_d A_d \eta v\cos\beta  \nonumber \\
&&
m_\tau= \lambda_\tau A_e v\cos\beta, ~~~~~~
m_\mu= \lambda_\mu A_e v\cos\beta, ~~~~~~~ 
m_e= \lambda_e A_e v\cos\beta
\eeqn
where the factors $A_f$ account for the gauge boson induced running 
from the scale $M_X$ down to the SUSY breaking scale $m_S\sim m_t$,  
$y$ accounts for running induced due to the 
large top Yukawa constant ($\lambda_t\sim 1$): 
\be{y}
y=\exp\left[-\frac{1}{16\pi^2}\int_{\ln m_t}^{\ln M_X}
\lambda_t^2(\mu)\mbox{d}(\ln \mu) \right]
\ee
and the factors $\eta_{b,c}$ (or $\eta$) encapsulate running from 
$m_t$ down to $\mu=m_{b,c}$ (or down to $\mu=1$ GeV for $u,d,s$). 
By taking $\alpha_3(M_Z)=0.11-0.13$, we have:
\beqn{RG_factors}
&&
\eta_b=1.5-1.6,~~~\eta_c=1.8-2.3,~~~
\eta =2.1-2.8, \nonumber \\
&&
A_u=3.3-3.8, ~~~ A_d=3.2-3.7,~~~ A_e=1.5
\eeqn
The RG running for neutrino masses was studied in 
refs.\cite{SV}. 

It is of obvious interest to find a self-consistent, complete and 
elegant enough example of a SUSY GUT that would provide a 
{\em realistic} and {\em predictive} framework for fermion mass and 
mixing pattern, and thus could be regarded as a Grand Unification 
of fermion masses. The naive concept of SUSY GUT solely is not 
sufficient to achieve this goal, 
and it should be complemented by other ideas that could  
further restrict the theory and thus enhance the predictivity.

\subsection{Grand DT Hierarchy Problem and small $\mu$ Problem }
\vspace{-0.35cm}

A realistic SUSY GUT should be capable to solve naturally the gauge 
hierarchy problem. At the level of the SM this is essentially a 
problem of the Higgs mass stability against radiative corrections 
(quadratic divergences). 
It is removed as soon as one appeals to SUSY, which links the 
scalar masses to those of their fermion superpartners while 
the latter are protected by the chiral symmetry. 
In the context of grand unification the gauge hierarchy 
problem concerns rather the origin of scales: why the weak scale 
$v\sim M_W$ is so small as compared to the GUT scale 
$M_X$, which in itself is not far from the Planck scale $M_P$.  
This question is inevitably connected with the 
doublet-triplet (DT) splitting puzzle\cite{Maiani}: the 
Higgs doublets $\phi_1,\phi_2$ embedded in the GUT multiplets are 
unavoidably accompanied by the colour triplet partners $\bar T,T$. 
The latter would mediate unacceptably fast proton decay (especially 
via $d=5$ operators\cite{dim5}) unless their masses are $\sim M_X$.  

For example, The Higgs sector of the minimal SUSY $SU(5)$ model 
consists of chiral superfields in adjoint (24 dimensional) 
representation $\Sigma$  and fundamental ($5 + \bar 5$) representations  
\be{HH}
H=(T+\phi_2), ~~~~~~~~~\bar H = (\bar T + \phi_1) 
\ee 
with $\phi_{1,2}$ being the MSSM Higgs doublets and $\bar T,T$ 
colour triplets. Superpotential involving these fields has a form: 
\be{W-su5}
{\cal W}=\frac{M}{2} \Sigma^2 + \frac{h}{3} \Sigma^3 + 
M_H \bar H H  +  f \bar H \Sigma H 
\ee 
The $SU(5)$ symmetry breaking down to 
$SU(3)\times SU(2)\times U(1)$ is provided by supersymmetric 
ground state $\langle \Sigma \rangle = (M/h)\mbox{diag}(2,2,2,-3,-3)$, 
$\langle \bar H, H \rangle =0$. Then the masses of the $T$ and $\phi$ 
superfields are respectively $M_3=M_H -(3f/h)M$ and $\mu=M_H + (2f/h)M$. 
So, the light doublet ($\mu\sim M_W$) versus heavy triplet 
($M_3\sim M_X$) requires  that $hM_H\approx -2fM$, with the accuracy 
of about $10^{-14}$.  Supersymmetry renders this constraint stable 
against radiative corrections.  
However, such a ``technical solution''\cite{Maiani} is nothing but 
the {\em Fine Tuning} of parameters in the superpotential. 

Natural solution of the DT problem can be provided by the 
``missing multiplet'' mechanism\cite{missing}. In this scenario the 
24-plet $\Sigma$ is substituted by the 75-plet $\Phi$ of $SU(5)$, 
and additional heavy superfields $\Psi+ \bar\Psi$ ($=50+\ov{50}$) are 
introduced. The latter have a bare mass 
term $M_\Psi \bar\Psi \Psi$,  $M_\Psi\sim M_X$, while $M_H H\bar H$ 
can be also suppressed by some symmetry. 
 In this case the term $\bar H \Phi H$ is absend, but instead 
the superpotential contains the terms $\bar H \Phi \Psi$ 
and $H \Phi \bar\Psi$. Then the triplet components $T,\bar T$ 
in $H,\bar H$ can get $\sim M_X$ mass via their mixing to 
triplet states in $\Psi,\bar\Psi$, while the doublets $\phi_{1,2}$, 
will remain massless since 50-plet does not contain doublet fragment. 

Yet another problem is a so-called $\mu$-problem\cite{mu}. 
The supersymmetric mass term $\mu\phi_1\phi_2$ should be small,  
and one could think that $\mu=0$ 
is a most natural possibility: then the Higgs masses would  
emerge entirely from the soft SUSY breaking terms with $m_S\sim v$. 
However, this is excluded experimentally, 
and $\mu$ is required to be  $\sim v$ as well.  
Hence, in SUSY GUTs the gauge hierarchy problem 
turns into a problem of small (but {\em not} vanishingly small) 
$\mu$-term: $\mu\sim m_S$. 
A realistic SUSY GUT should produce the supersymmetric 
$\mu$-term in a natural way, without fine tuning of $\mu$ to the 
soft SUSY breaking mass $m_S$.  

\subsection{Minimal $SU(5)$ Unification} 
\vspace{-0.35cm}

Already the minimal $SU(5)$ model can provide an important key  
towards understanding the fermion mass pattern. 
The quarks and leptons of each family fit into the multiplets 
\be{su5}
\bar{5}_i=(d^c_i + l_i), ~~~~~ 10_i=(u^c_i + q_i + e^c_i);  
~~~~~~~~~~ i=1,2,3 
\ee 
and the superpotential terms relevant for fermion masses become 
\be{su5-Yuk} 
{\cal W}_{\rm Yuk}= 
\la^u_{ij} 10_i H 10_j\, + \, \la^d_{ij} 10_i \bar{H} \bar{5}_j \,, 
~~~~~~~~~ 
{\cal W}_\nu= \frac{\la^{\nu}_{ij}}{M_P} (\bar 5_i H)(H \bar 5_j)
\ee 
At the GUT scale $M_X$ these terms reduce to the MSSM couplings 
(\ref{Y-MSSM}) with $\hat{\la}^e_{ij}=\hat{\la}^{d}_{ji}$, and hence 
$\la_{d,s,b}=\la_{e,\mu,\tau}$. 

The $\la_b=\la_\tau$ unification\cite{BEGN} is definitely a 
{\em Grand Prix}. After accounting for the RG running (\ref{RG}), 
it translates into $m_b/m_\tau=y\eta_bA_d/A_e\sim 3y$. 
When I get to the bottom I go back to the top\cite{Beatles}:
for $\la_t\ll 1$ ($y=1$) this explains basic factor of 3 
difference between the masses of bottom and tau.  However, 
the more precise comparison of $m_b$ and $m_\tau$ within the 
uncertainties in (\ref{masses}) points that $y<1$. This 
in turn implies a rather large $\la_t$ ($\geq 1$), in which case 
the top mass is fixed by its infrared limit\cite{IRfixed}
\be{Top}
M_t= (190-210)\sin\beta~ {\rm GeV}= 140-210~ {\rm GeV}
\ee 
Thus, the minimal SUSY $SU(5)$ model explains the principal origin 
of the bottom quark mass and nicely links it to the large value of 
the top mass, within the range indicated by the present data\cite{CDF}. 

Unfortunately, the other predictions 
$\la_s=\la_\mu$ and $\la_d=\la_e$ are wrong: they imply 
$m_s/m_d=m_\mu/m_e\simeq 200$ -- {\em c'est la vie!}
In addition, there is no explanation neither for the 
fermion mass hierarchy nor for the CKM mixing pattern: 
the Yukawa matrices $\la^u$ and $\la^d$ remain arbitrary 
and there is no reason for their allignment.  
Therefore, one is forced to go beyond the minimal $SU(5)$ model and 
implement new ideas that could shed some more light on the 
origin of fermion masses and mixing.

\subsection{Tools for Fermion Mass Models: Oldies but Goldies }
\vspace{-0.35cm}

Below we briefly review some ideas that can be regarded as a 
{\em Modus Operandi} for the predictive model building. 

$\bullet$ {\em Mass matrix textures}. 
Relations between the fermion masses and CKM angles can be obtained 
by considering the mass matrix ansatzes with reduced number of free 
parameters. In particular, certain elements in the Yukawa constant 
matrices can be put to zero (so called 
``zero textures")\cite{texture,Fritzsch}. One of the most popular 
ansatzes was suggested by Fritzsch\cite{Fritzsch}:
\be{Fr}
\hat{\la}_f=\,\matr{0}{A_f}{0}{A_f'}{0}{B_f}{0}{B'_f}{C_f},
~~~~~~~~~~f=u,d,e
\ee 
It implies that the fermion mass generation starts from the 
$3^{rd}$ family ($C$ is assumed to be a largest entry in eq. 
(\ref{Fr})) and proceeds to lighter families through the mixing terms. 
One can assume further that 
\be{A-B}
|A'_f|=|A_f|, ~~~~~~~ |B'_f|=|B_f|.
\ee 
Then, if neglect the phase factors, the total number of parameters for 
each matrix $\hat{\la}_{u,d,e}$ is reduced to 3, i.e. just 
the number of the quark and lepton species. This allows to express 
the quark mixing angles in terms of their mass ratios: 
\be{Fr-mix}
s_{12}=
\left|\sqrt{\frac{\la_d}{\la_s}}-
e^{i\delta}\sqrt{\frac{\la_u}{\la_c}}\right|, ~~~~
s_{23}=\left|\sqrt{\frac{\la_s}{\la_b}}-
e^{i\kappa}\sqrt{\frac{\la_c}{\la_t}}\right|, 
~~~~~s_{13}=\sqrt{\frac{\la_u}{\la_c}}\, s_{23} 
\ee 
where $\delta$ is a CP-violating phase and $\kappa$ is some unknown 
phase. In particular, when $\delta\sim 1$, we obtain 
$s_{12}\approx \sqrt{m_d/m_s}$. Unfortunately, the value of $s_{23}$ 
in eq. (\ref{Fr-mix}) is not compatible with the large top mass, 
neither in ordinary nor in supersymmetric cases\cite{KFP}. 
There are however some simple modifications\cite{Lavoura} 
of the ansatz which could still agree to the experimental data.  
For example, if $|B'_d|=2|B_d|$, one obtains a consistent value 
for $s_{23}$ (see below, eq. (\ref{Fr-mixed})). 
A dedicated analysis of possible zero-textures can be found 
in ref.\cite{RRR}. 

$\bullet$ {\em Radiative mechanism}. 
The observed mass hierarchy of about $1-2$ orders 
of magnitude between neighbouring families makes attractive the idea 
that radiative corrections may be responsible for mass generation. 
Masses of the light fermions could arise as a radiative effect 
from the tree-level masses of heavy family\cite{rad1,rad2}.  
Namely, if due to some reasons only the $3^{rd}$ family fermions 
have tree-level masses, the $2^{nd}$ family masses emerge at the 1-loop 
level and the $1^{st}$ family becomes massive only at 2-loops, then 
the inter-family hierarchy will have a shape of eqs. 
(\ref{qh}),(\ref{lh}), with $\eps_f\sim (g_f^2/16\pi^2)$ and 
 $g_f$ being typical coupling constants of the order of 1. 

Radiative models\cite{rad1,rad2} provide 
a rather qualitative explanation to the fermion mass hierarchy, 
and generally fail in predictivity. In particular, 
they cannot reproduce zero textures for mass matrices. 
Moreover, it is very difficult to obtain a quantitatively 
correct picture (e.g. $s_{12}\simeq \sqrt{m_d/m_s}$ contradicts 
to perturbativity), and also to avoide dangerous flavour changing 
phenomena\cite{NOP}.  

A consistent and predictive readiative approach was suggested 
in\cite{rad3}, where fermion 
mass hierarchy is first radiatively generated in the `hidden' sector of 
the heavy vectorlike fermions and the transfered in an inverted way to 
the usual quarks and leptons by means of the ``universal seesaw" 
mechanism. However, these models cannot be valid in the SUSY GUT context:  
at scales much larger than $m_S$ 
the loop corrections will be suppressed by supersymmetry. 

Nevertheless, this mechanism leaves the following important message: 
the lighter fermion masses could be due to the higher order operators. 
In gauge theories the vanishing of certain mass terms at the 
tree-level can occur as a consequence of the representation content 
of the fields in the theory, or due to some inter-family symmetry. 
In radiative scenario these mass terms then can emerge 
in the effective action as operators of dimension 
$d>4$ (i.e. involving more than one scalar leg). 

Within SUSY frames one could think of some tree level 
mechanism that could generate the relevant effective operators 
with successively increasing dimension, and 
thus explain the observed mass hierarchy.

$\bullet$ {\em Higher order (non-renormalizable) operators} (HOP).
Indeed, there is no physical reason for suppressing the 
higher order non-renormalizable terms scaled by inverse power of 
the Planck scale $M_P$. In fact, we have already introduced 
one such a term in (\ref{su5-Yuk}) for the neutrino mass generation.  
In the minimal $SU(5)$ theory one can consider\cite{EG} 
the higher dimension terms involving the 24-plet $\Sigma$:  
\be{hop}
\frac{1}{M_P}\, 10\,\Sigma H 10 \, + \, 
\frac{1}{M_P}\, 10\,\Sigma \bar H \bar 5 , ~~~~~~
\frac{1}{M_P^2}\, 10\, \Sigma^2 H 10 \, + \, 
\frac{1}{M_P^2}\, 10 \,\Sigma^2 \bar H \bar 5 ,~~ \dots
\ee 
which can be relevant for the light fermion masses. 
Below the scale $\langle \Sigma \rangle\sim M_X$ 
they contribute to the Yukawa couplings (\ref{Y-MSSM}) with 
the magnitudes $\sim \eps_X$, $\eps^2_X$ etc., $\eps_X=M_X/M_P$. 
This suggests that maybe the renormalizable couplings 
(\ref{su5-Yuk}) fix only the third family masses, 
thus maintaining the $\la_b=\la_\tau$ unification, and masses of 
the lighter families emerge entirely from the HOPs like 
(\ref{hop}). In this case one can avoide the {\em wrong} predictions 
$\la_{d,s}=\la_{e,\mu}$ of the minimal $SU(5)$, 
since the tensor product $24\times 5$ contains 
also the effective 45-plet.

Addressing the idea of mass matrix textures 
and also modifying the Higgs content of the theory, 
one can built more realistic and predictive models.   
For example, consider a ``missing partner'' model 
with 75-plet $\Phi$ instead of the 24-plet $\Sigma$, including 
also a singlet $Y$ with VEV $V_Y$. Let us also 
assume that due to some (inter-family) symmetry $G_H$ only 
the third family masses emerge from the direct Yukawa terms, while 
other entries are induced through the HOPs, and the lowest order 
terms in the superpotential allowed by $G_H$ are the following: 
\beqn{Phi-ops}
&& {\cal W}_{\rm up}=  10_3 H 10_3 + 
\frac{1}{M_P}\, 10_2(\Phi H)_{45} 10_3  + 
\frac{Y}{M_P^2}\, 10_1 (\Phi H)_{45} 10_2    \nonumber \\ 
&& {\cal W}_{\rm down}=  10_3 \bar H \bar 5_3 + 
\frac{1}{M_P}\, 10_2 (\Phi \bar H)_{\ov{45}} \bar 5_2 + 
\frac{Y}{M_P^2}\, (10_1 \bar H \bar 5_2 - 10_2 \bar H \bar 5_1) 
\eeqn 
(the order 1 coupling constants in each term are omitted). 
Then the Yukawa constant matrices at GUT scale obtain 
the following pattern:  
\be{GJ}
\hat{\la}^u=\matr{0}{-C}{0}{C}{0}{B}{0}{-B}{A} ~~~
\hat{\la}^d=\matr{0}{-F e^{i\delta}}{0}{F e^{-i\delta}}{E}{0}{0}{0}{D} 
~~~ \hat{\la}^e=\matr{0}{-F}{0}{F}{-3E}{0}{0}{0}{D}
\ee
It is a key moment that the tensor product $75 \times 5$ can induce 
the fermion masses only via the 45 channel. This creates the relative 
Clebsch coefficient $-3$ between the (2,2) entries in $\hat{\la}^d$ 
and $\hat{\la}^e$. On the other hand, 
$10\times 10$ contains 45 only in antisymmetric tensor product, 
so that no diagonal entries can be induced in $\hat{\la}^u$ while 
the (1,2) and (2,3) entries are antisymmetric. Interestingly, 
this also implies that quark couplings to the triplet component 
$T$ in $H$, $q_i q_j T$, 
being symmetric cannot emerge from these antisymmetric couplings. 
This will lead to partial suppression of the dangerous $d=5$ 
operators for the proton decay (see below, in setion 5).  

Except the (1,2) entry in $\hat{\la}^d$, all matrix elements in 
(\ref{GJ}) can be made real by proper phase redefinition of quark and 
lepton fields. Hence one is left with 8 parameters 
($A,B,C,D,E,F,\delta$ and $\tan\beta$) versus 14 observables 
(9 fermion masses, 3 mixing angles, CP-phase and $\tan\beta$), 
and thus 6 predictions can be obtained.

The ansatz (\ref{GJ}) was initially suggested by Georgi and 
Jarlskog\cite{GJ} in ordinary 
$SU(5)$ with the tree level Yukawa couplings involving the Higgs 5- 
and 45-plets. Its predictions in the supersymmetric 
case were elaborated at length in ref.\cite{DHR}. From (\ref{GJ}) 
one obtains (at the scale $\mu\sim M_X$) the Yukawa constant relations 
\be{GJ-mass}
\la_b=\la_\tau, ~~~~3(\la_s-\la_d) = \la_\mu -\la_e , 
~~~~ \la_d \la_s= \la_e \la_\mu
\ee 
and the following expressions for the CKM mixing angles 
\be{GJ-mix}
s_{12}= \left|\sqrt{\frac{\la_d}{\la_s}}-
e^{i\delta}\sqrt{\frac{\la_u}{\la_c}}\right|, ~~~~~
s_{23}=\sqrt{\frac{\la_c}{\la_t}}, 
~~~~~s_{13}=\sqrt{\frac{\la_u}{\la_c}}\, s_{23} 
\ee 
After accounting for the RG running, these predictions can be 
tested for the low energy (experimental) observables.  
We know that the Yukawa unification $\la_b=\la_\tau$ 
motivates the large value of $\la_t$. 
This in turn can render the value of $s_{23}$ compatible with 
experimental data, provided that $\sin\beta\sim 1$.\cite{DHR} 
For the light quark masses one obtains 
$m_s\sim 150$ MeV and $m_d/m_s=9m_e/m_\mu\approx 1/22$, in agreement 
with the current estimates (\ref{m-ratios}). In addition, for 
$\delta\sim 1$ we have $s_{12}\approx \sqrt{m_d/m_s}\approx 0.2$. 

Besides reproducing the predictive power of the ansatz\cite{GJ,DHR}, 
the operator structure in (\ref{Phi-ops}) can explain the origin of 
the fermion mass hierarchy. Indeed, we have: 
\beqn{GJ-eps}
&&
\la_t : \sqrt{\la_c\la_t}: \sqrt{\la_u\la_c} =  A:B:C \sim 
1 : \eps_X : \eps^2_X\eps_Y  \nonumber \\ 
&& ~~~
\la_\tau : \la_\mu : \sqrt{\la_e\la_\mu} = D:E:F \sim 
1 : \eps_X : \eps_X\eps_Y  
\eeqn 
where $\eps_X= M_X/M_P$ and $\eps_Y= V_Y/M_X$. The fermion mass 
pattern can be reproduced if $\eps_{X,Y}\sim 1/15$ or so; we have 
$\la_t/\la_c\sim \la_c/\la_u\sim \la_\mu/\la_e \sim 200$, as in eqs. 
(\ref{qh}) and (\ref{lh}). Then from $M_X\simeq 2\cdot 10^{16}$ GeV  
we obtain $M_P\sim 3\cdot 10^{17}$ GeV (string scale?), 
and $V_Y\sim 10^{15}$ GeV. 

$\bullet$ {\em Heavy fermion exchanges} (HFE).    
Higher order operators can be induced through 
the renormalizable interactions, as a result of integrating out the 
hypothetical superheavy particles\cite{FN,ZB1,ZB2,D}.
In other words, the quark and lepton masses can be induced through 
their mixings with the superheavy fermions, in a direct 
analogy to the celebrated seesaw mechanism for neutrinos\cite{seesaw}. 
Let us recall that in this scenario the RH neutrino $\nu^c$ is 
essentially a heavy neutral fermion with a Majorana mass $M\gg v$. 
The mass of physical (LH) neutrino emerges via its mixing term 
(=Dirac mass term) to the RH neutrino. Namely,  at scales below $M$ 
a diagram mediated by exchange of $\nu^c$ 
reduces to the effective operator (\ref{Ynu-SM}). 

One can introduce also a vector-like set of {\em charged} heavy fermions, 
having the same quantum numbers as the usual quark and lepton species. 
In the following we refer them to as $F$-fermions. 
Their exchanges can induce effective HOPs 
cutoff by a scale $M \sim M_F$ (not necessarily $M_P$). 
For example, the operators like (\ref{hop}) or (\ref{Phi-ops}) 
can be induced by $F$-fermions in representations\cite{ZB1,ZB2} 
\be{XV}
{\rm X}_k +\bar{\rm X}_k, ~~~~ \bar{\rm V}_l + {\rm V}_l, ~~~ {\rm etc.} 
\ee 
(following ref.\cite{ZB2}, we use roman numerals 
${\rm V}=5$ and ${\rm X}=10$ to denote their dimensions). 
Such fermions with mass $\sim M_P$ can exist in string theories.  
Typically they emerge also in the context GUT theories larger than 
$SU(5)$. E.g., 27-plet of the $E_6$ theory, in addition to $\bar 5+10$ 
contains also extra $\bar{\rm V} + {\rm V}$ states. 
In $SU(11)$ model\cite{surv}, after the $SU(11)$ symmetry breaking 
down to $SU(5)$, a number of $F$-states (\ref{XV}) 
emerge along with three chiral families of $\bar 5 + 10$.  
In spirit of the {\em survival hypothesis}\cite{surv}, they should 
get masses at the scales of the larger GUT symmetry breaking 
to $SU(5)$. Hence, generically the $F$-fermion  
masses and correspondingly cutoff scales of the effective HOPs,  
can vary from $M_P$ to $M_X$. 

The HOPs induced by the HFE mechanism can be more instructive  
for the fermion mass model building than the {\em ad hoc} 
introduced non-renormalizable operators. The HFE can produce 
the relevant operators in a rather selective way, 
and can fix the Clebsch factors between the Yukawa constants in 
dependence on the $F$-particles representations. 
Examples demonstrating advantages of the HFE mechanism will 
be given in next sections. 

In the literature the HFE mechanism, 
in its simplest form\cite{ZB1,unisaw} which is reminescent 
of the neutrino seesaw scheme, is also known as ``universal seesaw''. 
For its applications, see e.g. ref.\cite{Koide} and references 
therein. 

$\bullet$ {\em Horizontal (inter-family) symmetries.} 
The mass matrix textures can arise from to the spontaneously 
broken horizontal symmetry between the fermion families. 
 Consider, for example, model  with all quark and lepton states 
transforming as triplets $f_\al=(q,l,u^c,d^c,e^c)_\al$ 
of the hirizontal $SU(3)_H$ symmetry\cite{SU3H}, 
($\al=1,2,3$ is a family index). 
In order to avoide proliferation of Higgs doublets 
in non-singlet horizontal representations,  
the Higgs $\phi$ should be a singlet of the $SU(3)_H$.  
Extra light (with masses $\sim v$) Higgs doublets would spoil the 
natural suppression of the FCNC\cite{FCNC} and 
would also destroy the gauge coupling unification, thus preventing 
any attempt to embedd the model in SUSY GUT. 

Such a horizontal symmetry does not allow quarks and leptons 
to have renormalizable Yukawa couplings. Hence, the fermion mass 
generation is possible only after the $SU(3)_H$ breaking,  
through the HOPs involving some ``horizontal'' Higgses 
inducing this breaking at scales $V_H\gg v$.   
This suggests that observed mass hierarchy may emerge 
due to the hierarchy in the $SU(3)_H$ symmetry breaking.  

One can introduce horizontal scalars $\chi^{\al\beta}$ in 
the two-index symmetric or antisymmetric representations:   
say a sextet $\chi_3^{\{\al\beta\}}$ and two triplets 
$\chi_{1,2}^{[\al\beta]}\sim \eps^{\al\beta\ga}\chi_\ga$. 
Their VEV pattern can be chosen so that the sextet $\chi_3$ has a VEV 
$V_{33}$ towards (3,3) component, and triplets $\chi_2$ and $\chi_1$ 
have the smaller VEVs $V_{23}$ and $V_{12}$ directed towards 
1$^{st}$ and 3$^{rd}$ components. 
Thus, the total matrix of horizontal VEVs has a form\cite{ZB2}:  
\be{H-VEV} 
\hat{V}_H=\sum_k \langle \chi_k \rangle = 
\matr{0}{V_{12}}{0} {-V_{12}}{0}{V_{23}} {0}{-V_{23}}{V_{33}}, 
~~~~~~~~ V_{33}\gg V_{23}\gg V_{12} 
\ee
Then fermion masses can be induced by HOPs involving the 
horizontal Higgses $\chi_k$. For example, in the context of the 
$SU(5)\times SU(3)_H$ theory with fermions in representations  
$(\bar 5 + 10)_\al$. 
The relevant operators at lowest order are the following\cite{ZB2}: 
\be{su3h-su5}
{\cal W}_{\rm up}=
\frac{g^u_k\chi_k^{\al\beta}}{M} 10_\al H 10_\beta , ~~~~
{\cal W}_{\rm down}=
\frac{g^d_k\chi_k^{\al\beta}}{M}  10_\al \bar H \bar 5_\beta , 
~~~~~~~ {\cal W}_\nu = 
\frac{g^\nu_k \chi_k^{\al\beta}}{M^2} (\bar 5_\al H) (H\bar 5_\beta )
\ee 
These operators (\ref{su3h-su5}) can be induced through the HFE 
mechanism using the $F$-fermions ${\rm X}^\al + \bar{\rm X}_\al$ and 
$\bar{\rm V}^\al + {\rm V}_\al$,  
respectively in $(10 + \bar 5,\bar 3)$ and $(\ov{10}+5,3)$ 
representations of $SU(5)\times SU(3)_H$. 
Certainly, in this case the VEV pattern (\ref{H-VEV}) is 
directly reflected in the light fermion Yukawa matrices, 
and the fermion mass hierarchy follows to the 
$SU(3)_H$ symmetry breaking hierarchy. We refer this 
case as to a {\em direct hierarchy} pattern. 
In particular, the VEV pattern (\ref{H-VEV}) leads directly to 
the Fritzsch texture (\ref{Fr}) considered above.\footnote{
The $b-\tau$ Yukawa unification requires that $\chi_3$ is 
the $SU(5)$ singlet. As for triplets $\chi_{1,2}$, 
they should have a non-trivial $SU(5)$ assignment, say 24 or 75. 
Only in this case they can couple 10's in ${\cal W}_{\rm up}$ and 
thus generate antisymmetric off-diagonal entries in $\hat{\la}_u$  
(antisymmetric tensor product $10\times 10$ contains only the 
45-channel, which then can be confronted  by the effective 45  
in $24 \times 5$ product of the scalar fields).   
On the other hand, the effective $\ov{45}$ in ${\cal W}_{\rm down}$
will allow to split the down quark and charged lepton Yukawa 
constants in first two families. 
Alternatively, for purposes of economy, one could 
assume that $\chi_{1,2}$ are $SU(5)$ singlets,  
but by some symmetry reasoning they appear only in the next 
order terms together with $\Sigma$ or $\Phi$, like 
$(1/M^2)10\chi_{1,2}\Sigma H 10$, etc. }   

However, the HFE mechanism can suggest also another possibility, 
known as the  {\em inverse hierarchy} pattern\cite{ZB1}. Namely, 
the $F$-fermions can be introduced as $({\rm X} + \bar{\rm X})^\al$, 
etc. so that their invariant mass terms are forbidden by 
the $SU(3)_H$ symmetry. Then  they can get  masses after the 
horizontal symmetry breaking, 
via the Yukawa couplings $ {\rm X}\chi \bar{\rm X}$ etc.  
On the other hand, these fermions can mix the 
$(\bar 5 + 10)_\al$ states through the $SU(3)_H$ invariant 
couplings like $10 A \bar{\rm X}$, $ 10 H {\rm X}$, 
$\bar 5 \bar H {\rm X}$ etc., 
where $A$ is some dimensional parameter (or alternatively the 
Higgs 24 or 75 of $SU(5)$). Then, for $A\leq V_H$, the decoupling of 
the heavy states leads to effective operators\cite{ZB1}
\be{su3h-inv}
{\cal W}_{\rm Yuk}=
 10_\al \frac{AH}{\langle \sum g^u_k\chi^k_{\al\beta}\rangle} 
10_\beta \, + \,  
10_\al \frac{A\bar H}{\langle \sum g^d_k\chi^k_{\al\beta} \rangle} 
\bar 5_\beta 
\ee 
which {\em project} the VEV pattern (\ref{H-VEV}) on the fermion 
mass structure in the inverted way. The possible implications 
of the inverse hierarchy horizontal $SU(3)_H$ models were discussed 
in refs.\cite{SU3h,u1h}. 

The $SU(3)_H$ symmetry is attractive since it unifies all families. 
Within the same lines one can consider the models with a reduced 
horizontal symmetry $SU(2)_H$ acting bewteen first two families; 
then third family can get mass from the direct Yukawa couplings. 
Several other possibilities can be also envisaged, including discrete or 
abelian  horizontal symmetries\cite{LNS}.

$\bullet$ {\em P, CP, PQ and other flavour-blind symmetries}. 
Flavour-blind symmetries like spontaneoulsy broken 
P and CP parities  can also help in constraining the mass matrices. 
In particular, in the context of the $L\!-\!R$ symmetric 
model mass matrices should be Hermitian due to P-parity.  
For the Fritzsch ansatz (\ref{Fr}) this would imply a 
condition (\ref{A-B}). The spontaneously broken CP-invariance could 
constrain the complex phases of the Yukawa constants 
and thus enhance predictivity\cite{BCh}. 

P or CP  can provide a solution to the strong CP-problem 
without introducing an axion, {\em a l\'a} Nelson-Barr 
mechanism\cite{Nelson}. 
Such models based on ``universal seesaw'' were suggested in\cite{BM}, 
where the $\Theta$-term automatically vanishes at the tree level and 
emerges to be naturally small in the loop corrections. 
Alternatively, for the solution of the strong CP-problem 
one can introduce the Peccei-Quinn (PQ)\cite{PQ} type symmetries, 
which in addition could further restrict the mass matrix structure. 
In particular, in the horizontal $SU(3)_H$ symmetry models
the PQ symmetry can be naturally related to the phase 
transformation of the horizontal scalars $\chi$.\cite{ZB1,ZB2,SU3h}   
In this case axion appears to be simultaneously a majoron and familon. 

We conclude this section by demonstrating a $SU(5)\times SU(3)_H$ model 
which allows to properly correct the Fritzsch texture maintaining its 
predictive power. Let us introduce, along with the already familiar 
horizontal sextet $\chi_3$ and triplets $\chi_{1,2}$, also the 
$SU(3)_H$ octet scalar $\Lambda^\al_\beta$ with the VEV $\sim$ 
diag$(1,1,-2)$ towards the $\lambda_8$ direction. 
Let us also assume a `flavour-blind' discrete symmetry $Z_2$, under 
which the $\Lambda$ and $\bar 5_\al$ states change the sign  
while all other states are invariant. 
Then operator ${\cal W}_{\rm up}$ in (\ref{su3h-su5}) 
is still effective for the up quark masses, but 
${\cal W}_{\rm down}$ is forbidden now by the $Z_2$ symmetry. 
However, it can be replaced by the next order operator:
\be{su3h-la}
{\cal W}_{\rm down}=
\frac{g^d_k}{M^2} 
10_\al (\Lambda \chi_k)^{\al\beta}\bar H \bar 5_\beta 
\ee 
which can be obtained through the `double' exchange mediated by 
the $F$-fermions $\bar{\rm V}^\al + {\rm V}_\al$ ($Z_2$ singlets) 
and  $\bar{\rm V}'^\al + {\rm V}'_{\al}$ (wich also change the 
sign under $Z_2$). Then, by taking into account the VEV pattern 
(\ref{H-VEV}), we arrive to the following textures: 
\be{Fr-new}
\hat{\la}^u=\matr{0}{A_u}{0}{-A_u}{0}{B_u}{0}{-B_u}{C_u} ~~
\hat{\la}^d=\matr{0}{A_d}{0}{-A_d}{0}{B_d}{0}{2B_d}{C_d} ~~~
\hat{\la}^e=\matr{0}{-A_e}{0}{A_e}{0}{2B_e}{0}{B_e}{C_d} 
\ee 
Thus, the condition $|B_d|=|B'_d|$ is avoided 
and instead of eq. (\ref{Fr-mix}) we obtain 
\be{Fr-mixed}  
s_{12}=
\left|\sqrt{\frac{\la_d}{\la_s}}-
e^{i\delta}\sqrt{\frac{\la_u}{\la_c}}\right|, ~~~~
s_{23}=\left|\sqrt{\frac{\la_s}{2\la_b}}-
e^{i\kappa}\sqrt{\frac{\la_c}{\la_t}}\right|, 
~~~~~s_{13}=\sqrt{\frac{\la_u}{\la_c}}\, s_{23} 
\ee 
(notice the factor 2 in expression for $s_{23}$). 
Thus all predictions but for $s_{23}$ are the same as in the 
Fritzsch ansatz, and now the prediction for $s_{23}$ 
can perfectly fit its experimental value (\ref{angles}). 

In Section 4 we will demonstrate that flavour-blind symmetries 
can do much better job without any horizontal symmetry.

\section{Fermion masses in SUSY SO(10) } 
\vspace{-0.7cm}
\subsection{SO(10) Unification} 
\vspace{-0.35cm}

$SO(10)$ is a smallest group in which all the fermions in one 
family fit into one irreducible representation $16$. In 
addition to the quark and lepton states of the SM, it 
includes also RH neutrino $\nu^c$ (singlet of $SU(5)$). 
Thus, $SO(10)$ can in principle relate all Yukawa matrices 
$\hat{\la}^{u,d,e,\nu}$ by the $SO(10)$ Clebsch factors 
and thus reduce the number of fundamental parameters in the 
fermion sector. 

The $SO(10)$ symmetry can break down to $SU(3)\times SU(2)\times U(1)$ 
via two interesting channels: $SO(10)\to SU(5)$ or 
$SO(10)\to G_{422}= SU(4)\times SU(2)\times SU(2)'$. For the symmetry 
breaking purposes one has to introduce a set of Higgses in 
representations 54, 45 and $16+\ov{16}$, which we recall 
later on as $S$, $A$ and $\psi +\bar \psi$ fields. Their contents  
in terms of the $SU(5)$ and $G_{422}$ subgroups are the following: 
\beqn{54plet}
SU(5):  && 54=24+15+\ov{15}, ~~~~ 45=1+24+10+\ov{10}, 
   ~~~~       16=1 + \bar 5 + 10  \nonumber \\ 
G_{422}: && 54=(1,1,1)+(3,3,1)+(1,1,20')+(2,2,6), ~~~ 
16= (4,2,1) + (\bar 4,1,2), \nonumber \\ 
&&   45=(3,1,1)+(1,3,1)+(1,1,15)+(2,2,6)  
\eeqn 
The VEVs of $\psi,\bar\psi$ towards the $SU(5)$ singlet 
component reduces $SO(10)$ to $SU(5)$, while the VEV of $S$ contains 
only the $G_{422}$ singlet. The $SO(10)$ symmetry does not allow 
$S$ and $\psi,\bar\psi$ superfields 
to have renormalizable couplings to each other in superpotential. 
However, $A$ fields can couple both $S$ and $\psi$. 
One can introduce the Higgs 45-plets of the following types: 
``simple'' fields 
$A_{BL}$ having the VEV $V_{BL}$ only towards the (15,1,1) fragment, 
$A_R$ with VEV $V_R$ on the (1,1,3) fragment, 
$A_X$ with VEV $V_X$ towards the $SU(5)$ singlet component and  
$A_Y$ with VEV $V_Y$ orthogonal to $V_X$, and a ``general'' one 
$\tilde{A}$ having a VEV $\tilde{V}$ shared by both (15,1,1) and (1,1,3) 
components. The VEV orientation of the 45-plets are 
determined by their couplings to the $S$ and $\psi$ type superfields 
(see e.g. ref.\cite{BabuBarr}). In particular, $\tilde{A}$ has VEVs 
towards both (15,1,1) and (1,1,3) fragments if superpotential includes 
the both terms $\tilde{A}^2 S$ and $\psi\tilde{A}\bar\psi$.  
As for $A_{BL}$ and $A_R$, in order to ensure the strict `zeroes' in 
their VEVs, they should couple only to $S$ but not to $\psi\bar\psi$. 
On the contrary, $A_X$ should have no coupling with $S$ 
but only $\psi A_X\bar\psi$.
The trilinear terms like $A_{BL} A_R A_X$ are also necessary 
in order to evade the unwanted Goldstone modes. 

As for the Higgs doublets $\phi_{1,2}$, they fit into the 10-plet $H$ 
of $SO(10)$: 
\beqn{10plet} 
SU(5): ~~~ && 10=\bar 5(\bar T,\phi_1) + 5(T,\phi_2), \nonumber \\ 
G_{422}:~~~ && 10=\phi(1,2,2) + {\cal T}(6,1,1) 
\eeqn 
while three fermion families are arranged in chiral superfields 
$16_i$ ($i=1,2,3$):  
\beqn{16plet} 
SU(5): ~~~ && 
16_i=\bar 5(d^c,l)_i + 10(u^c,q,e^c)_i + 1(\nu^c)_i , \nonumber \\ 
G_{422}:~~~ && 16_i=f_i(4,2,1) + f_i^c(\bar 4,1,2) 
\eeqn 
For the fermion mass generation via the HFE mechanism 
one can also introduce superheavy families 
$16^F_k+\overline{16}^F_k$ ($k=1,2,...$): 
\beqn{16F}
SU(5): && 16^F_k= ({\rm X} + \bar{\rm  V} + {\rm I})_k, 
~~~~~~~~~~~~~
\ov{16}^F_k= (\bar{\rm X} + {\rm V} +\bar{\rm I})_k  \nonumber \\ 
G_{422}:  && 16^F_k\!=\!{\cal F}_k(4,2,1)+F^c_k(\bar{4},1,2), ~~~  
\ov{16}^F_k\!=\!{\cal F}^c_k(\bar{4},2,1)+F_k(4,1,2) 
\eeqn

In order to maintain the gauge coupling unification, we 
assume that all VEVs $V_S,V_\psi, V_{BL}$, etc. 
are of the order of $M_X\simeq 10^{16}\,$GeV, and below this scale 
SUSY $SO(10)$ theory reduces to the MSSM with three fermion families 
$f_i$ and the MSSM Higgses $\phi_{1,2}$  
(i.e. ${\cal X}(M_X)={\cal X}_{\rm MSSM}$). 
The field $A_{BL}$ can serve for the solution of the DT 
splitting problem through the ``missing VEV'' 
mechanism\cite{DiWi,BabuBarr}. In this way the Higgs doublets 
$\phi_{1,2}$ can be kept light while their colour triplet partners 
$\bar T,T$ acquire the $O(M_X)$ mass.

\subsection{Direct Hierarchy Approach: 
$t-b-\tau$ Unification and Large $\tan\beta$ } 
\vspace{-0.35cm}

One can assume that the masses of the heavy family emerge from a 
single renormalizable operator with the $O(1)$ coupling constant: 
\be{16-3}
O_{33}= 16_3\,H \,16_3 
\ee 
This leads to the unification of the Yukawa couplings 
$\la_t=\la_b=\la_\tau$, which is reminescent of the unification 
of the three gauge couplings $g_1$, $g_2$ and $g_3$.\cite{ALS}

The other fermion masses emerge from the higher order operators.  
In order to achieve predictivity, 
one has to appeal to the idea of zero textures. 
Recently various mass matrix ansatzes have been considered 
in the SUSY $SO(10)$ framework and several interesting (and testable) 
predictions were obtained\cite{ADHRS,BB}.  In particular, in 
ref.\cite{ADHRS} the so called ``22" textures were studied:  
\be{ADHRS} 
\hat{\la}^f= 
\matr{0}{z'_fC}{0}{z_fC}{y_fEe^{i\omega}}{x'_fB}{0}{x_fB}{A} , 
~~~~~~~f=u,d,e,(\nu) 
\ee 
where $z_f,z'_f$ etc. are the Clebsch factors distingusihing 
different fermions. Once the (3,3) entry emerges from (\ref{16-3}), 
its Clebsch is independent of $f=u,d,e$. 
Other Clebsches can be fixed by the structures of the higher order 
operators $O_{23}$, $O_{22}$ etc. generating  corresponding entries. 
In order to unambiguously fix these Clebsches, one has to accept 
the following rules of the game: (i) the relevant operators are 
obtained by means the HFE mechanism by exchange of heavy 
$16^F+\ov{16}^F$ states (\ref{16F}). These either have an invariant 
mass $M$ or acquire mass through the Yukawa couplings to the Higgs 
45-plets; 
(ii) all 45-plets involved into the game have `simple' VEV structure 
(i.e. there are only 45-plets $A_{BL,R,X,Y}$ 
and no `general' one $\tilde{A}$). 
 For example, in one of the models of ref.\cite{ADHRS} these operators 
have the following form, obtained by exchanges of 
9 heavy species $(16^F+\ov{16}^F)$: 
\be{HR-ops}
O_{23}=16_2 \frac{A_Y}{A_X} H \frac{A_Y}{A_X} 16_3, ~~~
O_{22}=  16_2 \frac{A_X}{M} H \frac{A_{BL}}{A_X} 16_2, ~~~
O_{12}= 16_1 \! \left(\frac{A_X}{M}\right)^3\!\! H\! 
\left(\frac{A_X}{M}\right)^3 \!\! 16_2 
\ee   
Then the fermion mass and mixing pattern is determined by 6 parameters 
$A,B,C,E,\omega$ and $\tan\beta$ which describe 14 observables. 
Thus 8 predictions can be obtained. A typical strategy for analysing 
these predictions is to fix six better known quantities: masses of 
the charged leptons ($m_e,m_\mu,m_\tau$), $c$,$b$ quark masses 
($m_c,m_b$) and the Cabibbo angle $s_{12}$ as input parameters, 
and deduce precise predictions for other, less known quantities: 
$m_{u,d,s}$, $M_t$, $s_{12}$, $s_{13}$ and CP-phase $\delta$.
In ref.\cite{ADHRS} a plethora 
of all possible operator structures was scanned, and it was shown 
that only very few of them could fit the observed pattern of 
the fermion masses and mixings. It is noteworthy that in all 
consistent cases the Clebsches $y_i$ satisfy $y_u:y_d:y_e=0:1:3$, 
the form familiar from the Georgi-Jarlskog ansatz\cite{GJ}. 
For the list of the testable predictions of the ansatzes 
(\ref{ADHRS}) one can address ref.\cite{ADHRS}.  

Needless to say, that one needs to invoke a very complex set of 
inter-family as well as flavour-blind symmetries which from one hand 
would enforce zeros in (\ref{ADHRS}) and, on the other hand, 
would fix the needed pattern of `non-zero' operators. 
These symmetries should also fix the Higgs 
superpotential terms in such a way that would force the 
Higgs 45-plets to have the VEVs of the desired `simple' pattern: only  
$X$, $Y$, $BL$ and $R$ directions, and also would implement 
the ``missing VEV'' mechanism\cite{DiWi} for the DT splitting. 

A key message of the {\em direct hierarchy} 
models\cite{ALS,ADHRS} is that third family plays a role 
of the Yukawa unification point at the GUT scale. 
It gets mass from the tree-level Yukawa term  
which fixes $\la_t=\la_b=\la_\tau$ at the GUT scale. Then 
mass generation proceeds to the lighter families through the smaller 
terms  induced by the HOPs. 
The large splitting of the top and bottom masses 
can be reconciled only at the price of extremely large tan$\beta$, 
of about two orders of magnitude. This can be achieved by certain 
tuning of parameters in the Higgs sector\cite{tanb}. 
However, it becomes then rather surprising that despite the giant 
tan$\beta$, $m_c/m_s$ is about 10 times less than 
$m_t/m_b$ while the first family is almost unsplit, 
$m_u \sim m_d$. Such a situation is reminescent of a fine tuning,  
and it deserves a judicious selection of the 
$SO(10)$ Clebsch coefficients. In particular, 
the operator $O_{12}$ in (\ref{HR-ops}), which implies 
$z_u : z_d : z_e = -\frac{1}{27} : 1 : 1$ ($z'_f=z_f$), 
appears to be the unique consistent one\cite{ADHRS}.

\subsection{Inverse Hierarchy Approach: 
$u-d-e$ Unification and Small $\tan\beta$ } 
\vspace{-0.35cm}

The fermion mass pattern (\ref{vert}), (\ref{vert-lept}) suggests 
that the first family might play a key role in understanding 
the structure of flavour: a role of the mass unification point.
Indeed, running masses of the first family exhibit an approximate 
$SO(10)$ symmetry:  $m_e\sim m_u\sim m_d$ with splitting of about a 
factor of 2, while the heavier families strongly violate it. 
In the context of small tan$\beta$ this may indicate that the 
$SO(10)$ Yukawa unification holds for the constants 
$\lambda_{u,d,e}$ rather than for $\lambda_{t,b,\tau}$. 

Such a possibility can be indeed realized by HFE mechanism 
in the {\em inverse hierarchy} approach\cite{ZB1}.  
The inter-family hierarchy could first emerge in a sector of 
$F$-fermions (\ref{16F}) and then be transfered inversely 
to ordinary quarks and leptons by means of the `universal seesaw'.  
With this picture in mind, it is suggestive to think that 
the $1^{st}$ family masses are unsplit since they are related to an 
energy scale $M_1> M_X$ at which the $SO(10)$ symmetry is still good, 
while the masses of the $2^{nd}$ and $3^{rd}$ family are respectively 
related to lower scales $M_{2,3}\leq M_X$, 
at which $SO(10)$ is no longer as good. 
 
A predictive inverse hierarchy framework in SUSY $SO(10)$ was 
suggested in ref.\cite{SO10}. This model involves the `simple' 
45-plets $A_{BL}$ with VEV $V_{BL}$ on the $(15,1,1)$ fragment, 
$A_R$ having a VEV $V_R$ towards the $(1,1,3)$ component, 
and a `general' one $\tilde{A}$ with the VEV $\tilde{V}$ shared 
by both (15,1,1) and (1,1,3) directions in terms of (\ref{54plet}).  
All these VEVs are assumed to be of the order of $M_X\sim 10^{16}$ GeV, 
and the only scale beyond is a string one $M_P=10^{17.5}$ GeV. 
The field $A_{BL}$ is used entirely for the DT splitting via 
the ``missing VEV'' mechanism, whereas $A_R$ and $\tilde{A}$ 
participate the fermion mass generation.  

It was assumed that $16_i$ (\ref{16plet}) have no 
direct Yukawa couplings $16_i H 16_j$ by certain (inter-family) 
symmetry reasons, and their masses {\em all} emerge through 
the `seesaw' mixing with three heavy families $(16^F+\ov{16}^F)_i$ 
(\ref{16F}). The relevant terms in superpotential are chosen as 
\be{fF} 
{\cal W}_{\rm mix}=\Gamma_{ij} 16_i H 16^F_j + 
\frac{G_{ij}}{M_P}\,(\overline{16}^F_i A_R)(A_R 16_j)  
\ee
(No concrete texture is specified for the coupling constants:  
$\hat{\Ga},\hat{G}$ are arbitrary non-degenarate matrices with 
$O(1)$ elements.) The heavy fermions themselves get masses 
from the operators of the successively increasing dimensions: 
\be{FF}
{\cal O}_{11} = M_P 16^F_1 \ov{16}^F_1,  ~~~~
{\cal O}_{22} = 16^F_2 \tilde{A}\, \ov{16}^F_2 , ~~~~  
{\cal O}_{33} = \frac{1}{M_P}\,(16^F_3 \tilde{A})(\tilde{A}\,\ov{16}^F_3)  
\ee  
${\cal O}_{33}$ as well as the second operator in (\ref{fF}) 
can be induced by the exchanges of 
some other heavy (with $M\sim M_P$) states $16^F+\ov{16}^F$, so 
that combinations in the parenthesses transform as 
effective 16 and $\ov{16}$.  

The choice of the $A_R$ type 45-plet in the mixing terms 
(\ref{fF}) is of key importance. In this case the light fermions 
in $16_i$ get mass only via the mixing to the weak 
isosinglet $F$-type components in the $16^F_i$ states (\ref{16F}), 
while the ${\cal F}$-type (weak isodoublet) fragments are irrelevant. 
This maintains the $SU(4)\times SU(2)\times SU(2)'$ invariance 
(i.e. quark-lepton and isotopic symmetries) in the `heavy-to-light' 
mixing terms and, interestingly, also suppresses the dangerous $d=5$ 
operators causing too fast decay of the proton\cite{PLB}. 

As a result, after integrating out the heavy states at the GUT scale, 
one obtains the Yukawa constant matrices in the following 
form:\footnote{
This pattern was first obtained in ref.\cite{rad3} 
in the context of the radiative mass generation scenario. }  
\be{mf-1}
\hat{\lambda}_{f}^{-1}= \frac{1}{\lambda}\, 
(\hat{P}_1+\eps_f\hat{P}_2+\eps_f^2\hat{P}_3 ) = 
\frac{1}{\la}
\matr{1+a^2\eps_f+x^2\eps_f^2}{ab\eps_f+xy\eps_f^2}{xz\eps_f^2} 
{ab\eps_f+xy\eps_f^2}{b^2\eps_f+y^2\eps_f^2}{yz\eps_f^2}
{xz\eps_f^2}{yz\eps_f^2}{z^2\eps_f^2}
\ee 
where $\la\sim (V_R/M_P)^2\sim 10^{-5}$, 
$\hat{P}_{1,2,3}$ are {\em rank-1} matrices with $O(1)$ elements, 
which without loss of generality are chosen as 
$\hat{P}_1\!=\!(1,0,0)^T\!\!\bullet(1,0,0)$, 
$\hat{P}_2\!=\!(a,b,0)^T\!\!\bullet(a,b,0)$, 
$\hat{P}_3\!=\!(x,y,z)^T\!\!\bullet(x,y,z)$,   
and $\eps_f\sim \tilde{V}/M_P\sim 10^{-1}-10^{-2}$ 
are small complex parameters, different for $f=u,d,e,(\nu)$.  
However, even if the VEV of $\tilde{A}$ has a `general' pattern, 
the $SO(10)$ symmetry imposes that  
$\eps_{d,u}=\eps_{15}\pm \eps_3$, $\eps_{e,\nu}=-3\eps_{15}\pm\eps_3$. 
Thus, only two of these four parameters are independent, and 
\be{eps2} 
\eps_e=-\eps_d-2\eps_u\,,~~~~~~~~ \eps_\nu=2\eps_e+3\eps_u 
\ee 

In lowest order the Yukawa constant eigenvalues are given 
by diagonal entries: 
$\lambda^{(i)}_{f}\approx\lambda\eps_f^{1-i}$ where $i=1,2,3$ is a 
family index. 
In this way, the quark mass pattern (\ref{vert}),(\ref{qh}) 
is understood by means of 
$\eps_u\ll\eps_d$: $~\lambda_{u,d}\sim\lambda$, 
$~\lambda_c/\lambda_s \sim (\eps_d/\eps_u)\sim 10$ and 
$\lambda_t/\lambda_b\sim (\eps_d/\eps_u)^2\sim 10^2$. This also 
implies that the CKM mixing emerges dominantly from the down 
quark matrix $\hat{\lambda}_d$: 
$\hat{\lambda}_u$ is much more "stretched" and essentially 
close to its diagonal form, so that it brings only $O(\eps_u/\eps_d)$ 
corrections to the mixing angles. Thus, at lowest order one expects 
that $s_{12},s_{23}\sim \eps_d$ and $s_{13}\sim\eps_d^2$, which 
estimates are indeed good for $s_{23}$ and $s_{13}$. 

However, for the correct quantitative picture 
the matrices (\ref{mf-1}) 
must be diagonalized\cite{SO10,rad3}  by assuming that  
$a^2\eps_f\sim 1$. Then for the Yukawa eigenvalues we obtain
\be{eigen} 
\la_{u,d,e}=\frac{\la}{|1+a^2\eps_{u,d,e}|}\,,~~~~~
\la_{c,s,\mu}=\frac{\la |1+a^2\eps_{u,d,e}|}{|b^2\eps_{u,d,e}|}\,,~~~~~
\la_{t,b,\tau}=\frac{\la}{|z^2\eps_{u,d,e}^2|} 
\ee 
while the CKM angles are 
\be{Cabibbo}
s_{12}=\frac{|\eps_dab|}{|1+\eps_da^2|}= 
\sqrt{\frac{\la_d}{\la_s}|\eps_da^2|}\,,~~~~
s_{23}=\frac{\lambda_u}{\lambda_d}\,\left|\frac{yz}{b^2}\eps_d\right|\,,
~~~~s_{13}=\frac{\lambda_d}{\lambda_u}\,|xz\eps_d^2|
\ee

From (\ref{eigen}) immediately follow the relations
\be{bottom1} 
\sqrt{\frac{\la_b}{\la_t}}=
\frac{\la_d\la_s}{\la_u\la_c}=\left|\frac{\eps_u}{\eps_d}\right|, 
~~~~~~~~~~\sqrt{\frac{\la_b}{\la_{\tau}}}=
\frac{\la_d\la_s}{\la_e\la_\mu}=\left|\frac{\eps_e}{\eps_d}\right|=
\left|1+2\,\frac{\eps_u}{\eps_d}\right|
\ee 
As soon as $|\eps_u/\eps_d|=\sqrt{\lambda_b/\lambda_t}\ll 1$, 
from (\ref{eps2}) we obtain that $\eps_d\approx -\eps_e$ 
and hence $\lambda_b\approx \lambda_\tau$ and 
$\lambda_d\lambda_s\approx \lambda_e\lambda_\mu$. 
Thus, in spite of naive expectation, the {\em Grand Prix} of $b-\tau$ 
unification is not lost: it is more precise the more $t-b$ are split. 
This is a direct result of the $SO(10)$ symmetry relation (\ref{eps2}). 
Otherwise, since $\la_{b,\tau}$ emerge at $O(\eps^2)$ level, e.g. 
the factor of 2 difference among $\eps_{d}$ and $\eps_{e}$ would cause 
already factor of 4 splitting between $\lambda_b$ and $\lambda_\tau$.  

On the other hand, the experimental value of the Cabibbo angle 
$s_{12}\simeq (m_d/m_s)^{1/2}$ implies that $|\eps_da^2|\simeq 1$. 
Then, owing to the same relation $\eps_d\approx -\eps_e$, 
the constants $\la_d=\la|1+\eps_da^2|^{-1}$ and 
$\la_e=\la|1+\eps_ea^2|^{-1}$  should deviate to different sides 
from $\la_u = \la$ by a factor of 2 or so. 
Hence, the first family can indeed play a role of the 
{\em Yukawa unification} point,  with its splitting 
understood by the same mechanism that enhances the Cabibbo angle 
up to the value $s_{12}\simeq\sqrt{m_d/m_s}$. 
The other mixing angles in (\ref{Cabibbo}) stay much smaller: 
$s_{23}\sim \eps_d$ and $s_{13}\sim \eps_d^2$, in accord to the 
observed pattern (\ref{mix-eps}).  

For more details of the model one can address refs.\cite{SO10}, 
where the detailed numerical analysis was carried out. 
In particular, the masses of leptons and $c$ 
and $b$ quarks, the ratio $m_s/m_d$ and the value of $s_{12}$ 
were taken as input parameters and 
the $u,d,s$ quark masses, top mass and $\tan\beta$ were computed:  
$m_s\simeq 150$ MeV, $m_d\simeq 7$ MeV and $m_u/m_d\simeq 0.5-0.7$. 
The top mass emerges in infrared-fixed regime ($\la_t\sim 1.5$ 
at GUT scale), and thus it is naturally in the 100 GeV range. 
However, there emerges an upper limit $\tan\beta <1.7$, which 
translates into the upper bound $M_t<165$ GeV. 

Zero textures of eq. (\ref{ADHRS}) give a bigger amount of predictions 
than the ansatz of eq. (\ref{mf-1}), as far as 
the model\cite{PLB}  is based on weaker assumptions than the models of 
ref.\cite{ADHRS}.  However, predictive power of the former 
approach can be enhanced further e.g. by imposing specific zero textures 
on the coupling constants in (\ref{fF}) which in general 
analysis were left arbitrary.

\vfil
\rm\baselineskip=14pt
\section{GIFT for Fermion Masses: SUSY SU(6) Model }
\vspace{-0.7cm}
\subsection{Higgs Doublets as Pseudo-Goldstone Bosons }
\vspace{-0.35cm}


SUSY $SU(6)$ model\cite{BD} (see also\cite{BDM,BDSBH,PLB,BCL}) 
was originally designed for the natural solution to the gauge hierarchy 
and doublet-triplet (DT) splitting problems via the elegant 
GIFT ({\em Goldstones Instead of Fine Tuning}) mechanism\footnote{The 
GIFT mechanism for the DT splitting was first suggested
in the context of SUSY $SU(5)$ by assuming an {\em ad hoc} $SU(6)$ 
global symmetry of the Higgs superpotential\cite{Inoue,Anselm}. 
Results for fermion masses, however, are specific of the gauged 
$SU(6)$ theory. }. 
The $SU(6)$ model\cite{BD} is a minimal extension of $SU(5)$:
the Higgs sector contains supermultiplets $\Sigma$ and
$H+\bar{H}$ respectively in adjoint 35 and fundamental $6+\bar{6}$
representations, in analogy to 24 and $5+\bar{5}$ of $SU(5)$.
However, this model drastically differs from the other GUTs 
where the Higgs sector usually consists of two different sets:
one is for the GUT symmetry breaking (e.g. 24-plet in $SU(5)$),
while another containing the Higgs doublets (like $5+\bar{5}$ 
in $SU(5)$) is just for the electroweak symmetry breaking. 
The $SU(6)$ theory has no special superfields for the second
purpose: 35 and $6+\bar 6$ constitute a minimal Higgs content
needed for the local $SU(6)$ symmetry breaking down to 
$SU(3)\times SU(2)\times U(1)$.
As for the MSSM Higgs doublets $\phi_{1,2}$, they emerge from 
$\Sigma$ and $H,\bar{H}$, as Goldstone modes of the
accidental global symmetry $SU(6)_\Sigma \times U(6)_H$.
This global symmetry arises if mixing terms of the form $\bar{H}\Sigma H$ 
are suppressed in the Higgs superpotential\cite{BD}. Then 
$\phi_{1,2}$ being strictly massless in the exact SUSY limit, acquire
non-zero mass terms (interestingly, including also the $\mu$-term) 
due to the spontaneous SUSY breaking. 

Once the Higgs doublets emerge as Goldstone modes, their couplings 
to fermions have peculiarities leading to new possibilities 
towards understanding of the flavour structure. 
Indeed, should the Yukawa terms also respect the 
$SU(6)_\Sigma \times U(6)_H$ global symmetry, then $\phi_{1,2}$ 
being the Goldstone modes would have the {\em vanishing} Yukawa 
couplings to all fermions which remain massless after the GUT 
symmetry breaking down to the MSSM, that are ordinary 
quarks and leptons. Thus, the couplings relevant for fermion 
masses have to {\em explicitly} violate $SU(6)_\Sigma \times U(6)_H$. 
This constraint leads to striking possibilities to understand 
the fermion mass and mixing pattern even 
in completely `democratic' approach, without invoking the 
the horizontal symmetry arguments. In particular, it was shown in 
ref.\cite{BDSBH} that {\em only} the top quark can get $\sim 100$ GeV 
mass through the renormalizable Yukawa coupling, 
while other fermion masses can emerge only through the higher order 
operators and thus are suppressed by powers of the Planck scale $M_P$.

In order to built a consistent GIFT model, one has to find some valid
symmetry reasons to forbid the mixing terms like $\bar{H}\Sigma H$:
otherwise the theory has no accidental global symmetry.
It is natural to use for this purpose the discrete symmetries,
which in principle could emerge in the string theory context.
In addition, they can provide a proper pattern of the higher 
order operators inducing the fermion masses. 
Possible consistent models were suggested in refs.\cite{PLB,BCL}. 
Below we consider the $SU(6)$ model of ref.\cite{PLB} with the 
flavour-blind discrete symmetry $Z_3$.

\subsection{ $SU(6)\times Z_3$ Model }
\vspace{-0.35cm}

We assume that below the Planck scale $M_P$ the field theory
is given by SUSY GUT with the $SU(6)$ gauge symmetry, 
equipped with the flavour-blind discrete symmetries $Z_3\times Z_2$. 
$Z_2$ stands for the usual matter parity, under which 
the fermion superfields change the sign while the Higgs 
ones stay invariant. $Z_2$ parity is known to be free of discrete 
anomalies\cite{Ibanez}, and is needed for suppressing the 
B and L violating $d=4$ operators. 
The theory contains the following chiral superfields: 

(i) Higgs sector: vectorlike set of supermultiplets $\Sigma_1(35)$,
$\Sigma_2(35)$, $H(6)$, $\bar{H}(\bar{6})$ and  
singlet $Y$;  

(ii) Fermion sector: chiral, anomaly free supermultiplets
$(\bar{6} + \bar{6}^\prime)_i$, $15_i$ ($i=1,2,3$ is a family index)
and 20;  

(iii) $F$-fermion sector: heavy vector-like matter multiplets like 
$15_F+\ov{15}_F$, etc.  with
large ($\sim M_P$) $SU(6)$ invariant mass terms. 
They will be needed for the light fermion mass
generation through the HFE mechanism\cite{FN,ZB1,ZB2,D}. 

The field content of the model and their $Z_3$ charges are given 
in Table 1. $Z_3$ symmetry satisfies the anomaly 
cancellation constraints\cite{Ibanez} and it can be regarded as 
a gauge discrete symmetry.

In the following we assume that all coupling constants in the Higgs 
as well as in the Yukawa sectors are of the order of 1. 
(For comparison, the gauge coupling constant
at the GUT scale is $g_X\simeq 0.7$.)  
The most general renormalizable Higgs superpotential compatible 
with the $SU(6)\times Z_3$ symmetry is 
\be{superpot}
{\cal W} = M_\Sigma\Sigma_1 \Sigma_2 + \lambda_1 \Sigma_1^3 + 
\lambda_2 \Sigma_2^3 + \lambda S \Sigma_1 \Sigma_2
+ M_H \bar{H}H + \rho Y (\bar{H}H - \Lambda^2) +
M_Y Y^2 + \xi Y^3
\ee
and it has an accidental global symmetry $SU(6)_{\Sigma}\times U(6)_H$, 
related to independent transformations  
of $\Sigma$ and $H$.\footnote{ Notice that global 
$SU(6)_\Sigma\times U(6)_H$ is not a symmetry of the whole Lagrangian: 
the Yukawa and the gauge couplings ($D$-terms) do not respect it.
However, in the exact supersymmetry limit it is effective for the 
field configurations on the vacuum valley where $D=0$. 
Owing to non-renormalization theorem, it cannot be
spoiled by the radiative corrections. }
In the limit of exact SUSY (i.e. of vanishing $F$ and $D$ terms), 
among the other degenerated vacua, there is the following one: 
\be{VEVs}
\langle \Sigma_{1,2} \rangle = V_{1,2} \left( \begin{array}{cccccc}
1 & & & & & \\& 1 & & & & \\ & & 1 & & & \\ & & & 1 & & \\ & & & & -2 & \\
& & & & & -2 \end{array} \right), ~~~~
\langle H \rangle =\langle \bar{H} \rangle = V_H \left( \begin{array}{c}
1\\0\\0\\0\\0\\0 \end{array} \right),~~~
\langle Y \rangle= V_Y
\ee
where, provided that $\Lambda\gg V_\Sigma=(V_1^2+V_2^2)^{\frac{1}{2}}$,
we have:
\be{VEV}
V_Y = \frac{M_H}{\rho},~~~~
V_{1,2}=\frac{ M_\Sigma + \lambda V_Y }
{ (\lambda_1\lambda_2)^{\frac{1}{3}}\lambda_{1,2}^{\frac{1}{3}}} \,,~~~~
V_H=\Lambda + O\left(\frac{V_\Sigma^2}{\Lambda}\right)
\ee
After SUSY breaking, the configuration (\ref{VEVs}) can indeed be 
a true vacuum state for a proper range of the soft parameters. Then 
$H,\bar{H}$ break $SU(6)$ down to $SU(5)$ while $\Sigma_{1,2}$ break
$SU(6)$ down to $SU(4)\times SU(2)\times U(1)$, and both 
channels together lead to the local symmetry breaking down to
$SU(3)\times SU(2)\times U(1)$. 
At the same time, the global symmetry $SU(6)_{\Sigma}\times U(6)_H$
is broken down to $[SU(4)\times SU(2)\times U(1)]_\Sigma \times U(5)_H$.
Most of the Goldstone degrees correspond to generators of the 
broken local $SU(6)$ and they are eaten up by the $SU(6)$
gauge superfields through the Higgs mechanism. However, since 
the global symmetry of the ground state exceeds the global one, 
a couple of fragments survive and present in particle spectrum at lower 
energies as the Goldstone superfields. These constitute 
the MSSM Higgs doublets $\phi_{1,2}$ which in terms of the doublet
(anti-doublet) fragments in $\Sigma_{1,2}$ and $H,\bar{H}$ are given as
\be{Higgs}
\phi_2= c_\eta(c_\sigma \phi_{\Sigma_1} + s_\sigma\phi_{\Sigma_2}) -
s_\eta \phi_H\,,~~~~~
\phi_1= c_\eta(c_\sigma \bar{\phi}_{\Sigma_1} +
s_\sigma\bar{\phi}_{\Sigma_2}) - s_\eta \bar{\phi}_{\bar{H}}
\ee
where $\tan\eta=3V_\Sigma/V_H$ and $\tan\sigma=V_2/V_1=
(\lambda_1/\lambda_2)^{\frac{1}{3}}\sim 1$. 
In the following  we assume that 
$M_P\gg V_H\gg V_\Sigma\simeq M_X \simeq 10^{16}$ GeV, 
as it is motivated by the $SU(5)$ unification of the gauge couplings.  
In this case the doublets $\phi_{1,2}$ 
dominantly come from $\Sigma_{1,2}$
while in $H,\bar{H}$ they are contained with small
weight $\sim 3V_\Sigma/V_H$.

\begin{table}[t]
\begin{center}\begin{tabular}{c|c|c|l}
$Z_3$: & Higgs & fermions & ~~~~ $F$-fermions \\ \hline
$\omega$ & $\Sigma_1$ & $\bar{6}_i$, $\bar{6}'_i$, $20$ & $\ov{15}^2_F$,~
$\ov{15}^3_F$,~ $20_F$,~ $35_F$,~ $\ov{70}_F$,~ $84_F$ \\ \hline
$\bar{\omega}$ & $\Sigma_2$ & $15_i$ & $15^2_F$,~ $15^3_F$,~
$\ov{20}_F$,~ $\ov{35}_F$,~ $70_F$,~ $\ov{84}_F$ \\ \hline
{\em inv.} & $H$, $\bar{H}$, $Y$ & -- & $\overline{15}^1_F$,~
$15^1_F$,~ $20_F^{1,2}$,
$\overline{105}_F$, $105_F$, $\overline{210}_F$, $210_F$
\end{tabular}

\caption{$Z_3$-transformations of various $SU(6)$ supermultiplets 
($\omega=\exp ({\rm i}2\pi/3)$ ). }
\end{center}\end{table}


The scalar fields in $\phi_{1,2}$ then get mass from 
the soft SUSY breaking terms\cite{BFS}: 
\be{SB_terms}
V_{SB}=Am_S{\cal W}_3 + Bm_S{\cal W}_2 + m_S^2\sum_k|\varphi_k|^2,
\ee
where $\varphi_k$ imply all scalar fields involved, ${\cal W}_{3,2}$
respectively are trilinear and bilinear terms in (\ref{superpot}) 
and $A,B,m_S$ are the soft breaking parameters.
SUSY breaking relaxes radiative corrections which lift the vacuum 
degeneracy (mainly due to the large top Yukawa coupling, origin of 
which we will clarify below) and fix the VEVs $v_1$ and $v_2$. 
The effects of radiative corrections leading to the electroweak 
symmetry breaking were studied recently in refs.\cite{CR}. 

Interestingly, the GIFT scenario naturally solves also the
$\mu$-problem. Taking into account the soft SUSY breaking terms 
(\ref{SB_terms}) in minimization of the Higgs potential of $\Sigma$ 
and $H,\bar H$, one observes that the VEVs 
$V_{1,2}$ are shifted by an amount of $\sim m_S$ as compared to 
the ones of eq. (\ref{VEV}) calculated in the exact SUSY limit. 
Then substituting these VEVs back in superpotential, this shift 
gives rise to the $\mu \phi_1 \phi_2$ term, with $\mu\sim m_S$. 
Thus, in GIFT scenario the (supersymmetric) $\mu$-term   
emerges as a consequence of the SUSY breaking.

Thus, the $SU(6)$ model naturally solves both the DT splitting and the
$\mu$ problems. The Higgs doublets $\phi_{1,2}$ remain light,
$\mu\sim m_S$, and their triplet partners are superheavy: 
the triplets from $\Sigma_{1,2}$ have masses
$\sim M_X$, while the `Goldstone' triplets from 
$H, \bar H$ acquire masses $\sim V_H$ due to the Higgs mechanism 
by mixing with the $SU(6)$ gauge superfields.

\subsection{ Fermion Masses and Mixing in $SU(6)\times Z_3$ Model }
\vspace{-0.35cm}

Now we show how the observed hierarchy of fermion masses 
and mixings can be naturally explained in terms of small ratios
$\eps_\Sigma=V_\Sigma/V_H$ and $\eps_H=V_H/M_P$. 

The most general Yukawa superpotential allowed by the
$SU(6)\times Z_3$ symmetry is
\be{Yukawa} 
{\cal W}_{Yuk} = G\,20 \Sigma_1 20 \,+ \,\Gamma\,20 H 15_3\,  + \,
\Gamma_{ij} 15_i \bar{H} \bar{6}^\prime_j\,, ~~~~~~~~i,j=1,2,3
\ee 
where all coupling constants are $O(1)$. Without loss of generality, 
one can always redefine the basis of 15-plets
so that only the $15_3$ state couples 20-plet in (\ref{Yukawa}).
Also, among six $\bar{6}$-plets one can always choose three of them
(denoted in eq. (\ref{Yukawa}) as $\bar{6}^\prime_{1,2,3}$) which couple
$15_{1,2,3}$ while the other three states $\bar{6}_{1,2,3}$ have
no Yukawa couplings.

Already at the scale $V_H$ of the gauge symmetry breaking
$SU(6) \to SU(5)$ the fermion content reduces to the one of the 
minimal $SU(5)$. Indeed, in terms of the $SU(5)$ subgroup the 
fermions under consideration read as 
\beqn{fragments}
& & 20=10 + \ov{10} = (q+u^c+e^c)_{10}+(Q^c+U+E)_{\ov{10}} \nonumber \\
& & 15_i=(10+5)_i = (q_i+u^c_i+e^c_i)_{10} +
(D_i + L^c_i)_5  \nonumber \\
& & \bar{6}_i=(\bar{5}+1)_i = (d^c_i + l_i)_{\bar{5}} + n_i \nonumber \\
& & \bar{6}_i^\prime=(\bar{5}+1)_i^\prime =
(D^c_i +L_i)_{\bar{5}^\prime} + n_i^\prime\,,
~~~~~~~~~~~~~~~~~~~~~i=1,2,3
\eeqn
In the spirit of survival hypothesis\cite{surv}, 
the extra vector-like fermions $\ov{10}+10_3$ and $(5+\bar{5}')_{1,2,3}$,
get $\sim V_H$ masses from couplings (\ref{Yukawa}):
\be{heavymass}
\Gamma\,V_H\,\ov{10}\,10_3\, +
\,\Gamma_{ij}V_H\,5_i\,\bar{5}_j^\prime\, +
\,G\, V_1 \,(U\,u^c - 2 E\,e^c) \,,
\ee
and thereby decouple from the light states which remain as
$\bar{5}_{1,2,3}$, $10_{1,2},10$ and singlets $n_i,n'_i$ 
(we neglect $\sim \eps_\Sigma$ mixing between the 
$u^c - u^c_3$ and $e^c - e^c_3$ states). 

The couplings of 20-plet in (\ref{Yukawa}) explicitly
violate the global $SU(6)_\Sigma\times U(6)_H$ symmetry. Hence,
the up-type quark from 20 (to be identified as top)
has {\em non-vanishing} coupling with the Higgs doublet $\phi_2$.
As far as $V_H\gg V_\Sigma$, it essentially emerges
from $G\,20\Sigma_1 20 \to G\, q u^c \phi_2.\,$
Thus, {\em only} the top quark can have $\sim 100\,$GeV
mass due to the large Yukawa constant $\lambda_t=G \sim 1$.

Other fermions would stay massless unless we invoke the HOPs 
scaled by inverse powers of $M_P$, 
which will emerge by integrating out the $F$-fermions.  
Before addressing the concrete HFE scheme, let us first analyse the 
general structure of the possible HOPs. 
Obviously, $Z_3$ symmetry forbids the $d=5$ `Yukawa' terms in the
superpotential. However, the following $d=6$ and $d=7$ operators are 
allowed:\footnote{
The terms like $15 \bar H (\Sigma_1\Sigma_2 \bar 6)$ or 
$15 \bar H \bar 6\cdot {\rm Tr}(\Sigma_1\Sigma_2)$ do not violate 
the global $SU(6)_\Sigma\times U(6)_H$ symmetry and are 
therefore irrelevant. }
\beqn{dim6}
d=6: ~~~~
&{\cal B} = \frac{B}{M_P^2}\, 20 \bar{H} (\Sigma_1 \bar{H}) \bar{6}_3\,,
~~~~{\cal C} = \frac{ C_{ij} }{M_P^2}\, 15_i H (\Sigma_2 H) 15_j
\nonumber \\ 
&{\cal S} = \frac{S^{(1)}_{ik} }{M_P^2}\, 15_i(\Sigma_1\Sigma_2\bar{H})
\bar{6}_k +
\frac{S^{(2)}_{ik} }{M_P^2}\,15_i(\Sigma_1 \bar{H})(\Sigma_2 \bar{6}_k)
\nonumber \\ 
&{\cal N} = \frac{N_{kl} }{M_P^2}\, \bar{6}_k H (\Sigma_1 H) \bar{6}_l
\eeqn
\beqn{dim7}
d=7: ~~~~
& {\cal D}= \frac{D^{(1)}_{ik}}{M_P^3} 15_i (\Sigma_{1}^3 \bar{H})\bar{6}_k
+ \frac{D^{(2)}_{ik}}{M_P^3} 15_i (\Sigma_{1}^2 \bar{H})
(\Sigma_1 \bar{6}_k) + \nonumber \\ 
&\frac{D^{(3)}_{ik}}{M_P^3} 15_i (\Sigma_{1} \bar{H})
(\Sigma_1^2 \bar{6}_k) +  \frac{D^{(4)}_{ik}}{M_P^3} 15_i 
(\Sigma_{1} \bar{H}) \bar{6}_k\cdot {\rm Tr}\Sigma_1^2 
\nonumber \\ 
& {\cal U} = \frac{U^{(1)}_{ij}}{M_P^3} 15_i H (\Sigma_1^2 H) 15_j +
\frac{U^{(2)}_{ij}}{M_P^3} 15_i H (\Sigma_1 H) \Sigma_1 15_j
\eeqn 
($SU(6)$ indices are always contractrd so that combinations 
in the parentheses transform as effective $\bar 6$ or $6$).
Since the $\bar 6'$ and $15_3$ states already have $\sim V_H$ masses, 
these operators are relevant only for the light states in 
$20$, $15_{1,2}$ and $\bar{6}_{1,2,3}$. 
One can always redefine the basis of $\bar{6}$-plets so that only 
the $\bar{6}_3$ couples 20 in eq. (\ref{dim6}). In addition, we 
assume that constants $B$, $C_{ij}$, etc. all are order 1 as well 
as the constants in (\ref{Yukawa}). 

Operator ${\cal B}$ gives rise to the $b$ quark
and $\tau$ lepton masses. At the MSSM level it reduces to the Yukawa
couplings $\eps_H^2 B (q d^c_3  + e^c l_3)\phi_1$.
Hence, though $b$ and $\tau$ belong to the 20-plet as well as 
$t$, their Yukawa constants are by factor $\sim \eps_H^2$ smaller 
than $\lambda_t$. In addition, the $b-\tau$ Yukawa constants are 
automatically unified at the GUT scale: $\la_b=\la_\tau$,  
up to $\sim \eps_\Sigma^2$ corrections due to 
the mixing of $e^c$ and $e^c_3$ states in eq. (\ref{heavymass}).

Operator ${\cal C}$ contributes the up quark Yukawa constants 
of the first and second families, as 
$\lambda_{ij}^u=\eps_H^2 C_{ij}$ ($i,j=1,2$). 
As for the operators ${\cal S}$ and ${\cal D}$, they induce the Yukawa 
constants of the down quarks and charged leptons respectively as 
\beqn{Yuk_de}
\lambda_{ik}^d=\eps_\Sigma\eps_H^2 
(S^{(1)}_{ik} - S^{(2)}_{ik}), ~ 
&&
\tilde{\lambda}^d_{ik} = \eps_\Sigma^2\eps_H^3 
(3D_{ik}^{(1)}- D_{ik}^{(2)} +D_{ik}^{(3)} +12D_{ik}^{(4)}) \nonumber \\
\lambda_{ik}^e=\eps_\Sigma\eps_H^2 
(S^{(1)}_{ik} + 2S^{(2)}_{ik}), ~ 
&&
\tilde{\lambda}^e_{ik} = \eps_\Sigma^2\eps_H^3 
(3D_{ik}^{(1)} + 2D_{ik}^{(2)} + 4D_{ik}^{(3)} + 12D_{ik}^{(4)})
\eeqn
($i=1,2$, $k=1,2,3$). Clearly, the mass hierarchy between the first 
and second families fermions can have a realistic shape only if 
the latter emerge from the $d=6$ operators ${\cal C}$ and ${\cal S}$, 
while the first family 
get masses from the $d=7$ operators ${\cal U}$ and ${\cal D}$.
Certainly, this can be done by introducing some horizontal 
symmetry which could fix the mass matrix textures and the structure 
of the HOPs involved in mass generation (see e.g. ref.\cite{BCL}). 
However, it is interesting that in the context of the HFE mechanism
the basic explanation of the fermion mass and mixing pattern can 
achieved in a completely ``democratic'' approach, without appealing 
to any horizontal symmetry. This can be obtained as a result of the 
properly chosen representations for $F$-fermions. 
In particular, the HFE can induce the HOPs (\ref{dim6}) in such a 
manner that $C_{ij}$ and $S^{1,2}_{ik}$ emerge as the 
{\em rank-1} matrices, which then 
without loss of generality can be chosen as 
\be{rank-1}
C_{ij}=\mat{0}{0}{0}{C}, ~~~~~~ 
S^{(1,2)}_{ik} \propto \left(\begin{array}{ccc} 
{0}&{s_\theta S_2}&{s_\theta S_3} \\
{0}&{c_\theta S_2}&{c_\theta S_3} \end{array}\right)
\ee
In addition, such HOPs should provide definite Clebsch structures,  
which would allow to obtain certain mass relations. 

The relevant HFE's involving the $F$-fermions of Table 1 
are shown in Figs. 1-3. As a result, one 
obtains the following pattern of the Yukawa couplings 
at the GUT scale $M_X$: 
\begin{eqnsystem}{sys:ude}
&&\bordermatrix{& u^c_1 & u^c_2 & ~u^c \cr
q_1 & 0 & \eps_\Sigma\eps_H^3 U & ~0 \cr
q_2 & \eps_\Sigma\eps_H^3 U' & \eps_H^2 C & ~0 \cr
q   & 0 & 0 & ~G \cr} \cdot \phi_2 \\
&&
\bordermatrix{& d^c_1&d^c_2 & d^c_3 \cr
q'_1  & J\eps_\Sigma^2\eps_H^3 D_{1}
      & J\eps_\Sigma^2\eps_H^3 D_{2}
      & J\eps_\Sigma^2\eps_H^3 D_{3} \cr
q'_2  & J\eps_\Sigma^2\eps_H^3 D'_{2}
      & K\eps_\Sigma\eps_H^2 S_2
      & K\eps_\Sigma\eps_H^2 S_3 \cr
q     & 0 & 0 & \eps_H^2 B \cr} \cdot   \phi_1 \\
&&
\bordermatrix{& l_1 & l_2 & l_3 \cr
e'^c_1 & \eps_\Sigma^2\eps_H^3 D_{1}
       & \eps_\Sigma^2\eps_H^3 D_{2}
       & \eps_\Sigma^2\eps_H^3 D_{3}  \cr
e'^c_2 & \eps_\Sigma^2\eps_H^3 D'_{2}
       & \eps_\Sigma\eps_H^2 S_2
       & \eps_\Sigma\eps_H^2 S_3 \cr
e^c    & 0 & 0 & \eps_H^2 B \cr} \cdot   \phi_1
\end{eqnsystem} 
Notice that the basis of down quarks (in $15'_{1,2}$) is already
`Cabibbo' rotated with respect to the upper quarks basis 
$15_{1,2}$, by the angle $\theta\sim 1$ (see eq. (\ref{rank-1})). 
The HFE shown in Figs. 1,3 induce operators ${\cal S}$ and 
${\cal D}$ respectively in conbinations 
${\cal S} \propto {\cal S}_1+2{\cal S}_2$ 
and ${\cal D} \propto {\cal D}_1+{\cal D}_3-{\cal D}_4$
(in terms of the possible operators in (\ref{dim6}) and (\ref{dim7})). 
Then the Clebsch factors are fixed as $J=8/5$ and $K=-1/5$. 

From (\ref{sys:ude}) we obtain 
the Yukawa coupling eigenvalues at the GUT scale 
\beqn{Yuk_pattern}
 3^{rd} ~{\rm family}:~~~~ & \lambda_t\sim 1,
&\lambda_\tau=\lambda_b\sim \eps_H^2     \nonumber \\
 2^{nd} ~{\rm family}:~~~~ & \lambda_c\sim \eps_H^2,  
&\lambda_\mu= 5\lambda_s \sim \eps_\Sigma\eps_H^2     \nonumber \\
 1^{st} ~{\rm family}:~~~~ & \lambda_u\sim \eps_\Sigma^2\eps_H^4, 
&\lambda_e= \frac{5}{8} \lambda_d \sim \eps_\Sigma^2\eps_H^3 
\eeqn  
while the CKM angles are (notice different parametrization 
from that of eq. (\ref{CKM})): 
\be{CKM-new}
V_{\rm CKM}\approx
\matr{ 1 }{ s_{12} }{ s_{12}s_{23} - s_{13}e^{-i\delta} }
{ -s_{12} }{ 1 }{ s_{23} + s_{12}s_{13}e^{-i\delta} }
{ s_{13}e^{i\delta} }{ -s_{23} }{ 1 }, ~~~~
s_{12}(\theta)\sim 1,~~
s_{23}\sim \frac{\lambda_s}{\lambda_b}, ~~
s_{13}\sim \frac{\lambda_d}{\lambda_b}
\ee
where the CP-phase $\delta$ arises due to the complex Yukawa 
constants in the theory.  

Taking into account the RG running for the Yukawa constants 
(\ref{RG}), this pattern can be confronted to the low energy 
(experimental) observables (\ref{masses}) and (\ref{angles}). 
We see that in the context of small $\tan\beta$ (which is rather 
natural in the GIFT scenario), eqs. (\ref{Yuk_pattern}) and 
(\ref{CKM-new}) explain all basic features 
of the fermion masses and mixings in terms of 
small parameters $\eps_H,\eps_\Sigma\sim 0.1$ (compare e.g. with 
pattern of eqs. (\ref{pattern}) and (\ref{mix-eps})). 
Moreover, from (\ref{Yuk_pattern}) follows that 
\be{d/s}
\frac{m_d}{m_s}\simeq 8\,\frac{m_e}{m_\mu}\approx \frac{1}{25}\,
[1 + O(\eps_\Sigma)]
\ee
($\sim \eps_\Sigma$
correction can arise due to mixing terms in (\ref{sys:ude})), 
while for the $s$ quark running mass at $\mu=1\,$GeV we obtain
$m_s = (\eta A_d/5A_e) m_\mu = 100-150$ MeV.

As for the neutrinos, they acquire small Majorana mass 
from the operator ${\cal N}$ 
in (\ref{dim6}) induced by the HFE in Fig. 2:  
\be{nu_mass}
m^\nu_{ij}\sim N_{ij}\frac{\eps_H v^2 }{\eps_\Sigma M_P}
\ee
which for $\eps_\Sigma,\eps_H\sim 0.1$ gives  
$m_\nu \sim v^2/M_P \sim 10^{-5}\,$eV. 
Since our `democratic' approach in general implies no hierarchy 
in constants  $N_{kl}\sim 1$, one expects that neutrino mixing 
angles are $O(1)$. Thus the predictions for the neutrino oscillation 
parameters are in the range (\ref{JS}), needed for the ``just-so'' 
solution to the solar neutrino problem\cite{just-so}. 

Let us remark that the above results are obtained from the 
general operator analysis of all possible HFE, 
and the $F$-fermion content of the Table 1 is uniquely selected 
among several other possibilities\cite{PLB}.
In constructing the HOPs the following constraints 
have been taken into account: 

(A) In order to ensure that 
the $d=6$ operators ${\cal C}$,${\cal S}$ induce only the 
second family fermion ($c,s,\mu$) masses, the have to be
induced by the unique exchange chain.

(B) Once the HFE generating ${\cal C}$ and ${\cal S}$ are
selected, the $d=7$ operators ${\cal D}$ and ${\cal U}$ should be
constructed by the $F$-fermion exchange chains which are
irreducible to $d=6$ operators: otherwise the mass hierarchy
between the first and second families would be spoiled.
In other words, the exchange chains should not allow to replace
$\Sigma_1\times\Sigma_1$ by $\Sigma_2$, so that the (symmetric) tensor
product $\Sigma_1\times\Sigma_1$ should effectively act as the
$189$ or $405$ representations of $SU(6)$.
This condition requires the large representations like 105, 210, etc.
to be involved into the game.

All possible HFE satisfying the conditions (A) and (B) have 
been classified\cite{PLB}. It was shown, that 
operator ${\cal D}$ can be induced only by few other irreducible
chains involving large representations, 
which give rise to the combinations 
${\cal D}_1-{\cal D}_2+{\cal D}_3+{\cal D}_4$: ($J=1$), 
${\cal D}_1\mp {\cal D}_4$: ($J=1$), 
${\cal D}_1-2{\cal D}_2-{\cal D}_4$: ($J=11/17$), and 
${\cal D}_1+{\cal D}_3-{\cal D}_4$: ($J=8/5$). 
Therefore, the HFE of Fig.~3  implying 
$J=8/5$ is selected as the only one feasible choice: 
all other possibilities lead to $\lambda_d\leq\lambda_e$ and 
are thus unacceptable. 

Also the HFE relevant for operator ${\cal S}$ have been classified.
By scanning all possible representations for the $F$-fermions,
it has been obtained\cite{PLB} 
that ${\cal S}$ can emerge only in the combinations
${\cal S}_1$: ($K=1$), ${\cal S}_2$: ($K=-1/2$),
${\cal S}_1\pm {\cal S}_2$: ($K=0,-2$ respectively),
${\cal S}_1-2{\cal S}_2$: ($K=-1$), and
${\cal S}_1+2{\cal S}_2$: ($K=-1/5$). We have chosen the latter case
uniquely selected by the HFE in Fig.~1.
All other cases are unacceptable: $K=0$ ($|K|\geq 1$) leads to massless
(too heavy) $s$ quark,
while $K=-1/2$ in combination with $J=8/5$ implies 
$m_d/m_s\approx 1/70$.

As for the operators ${\cal C}$ and ${\cal U}$, the only possible 
HFE obeying conditions (A) and (B) are the ones shown in Figs. 1,3.
Thus, among all possible $F$-fermions only the ones selected 
in Table 1 lead to acceptable pattern of the HOPs in (\ref{dim6}) 
and (\ref{dim7}). 

Concluding, the fermion mass and mixing pattern
can be naturally explained in our scheme without appealing to
any horizontal symmetry, provided that the scales $M_P$, $V_H$ and
$V_\Sigma$ are related as $V_\Sigma/V_H \sim V_H/M_P \sim 0.1$.
As far as  the scale
$V_\Sigma\simeq 10^{16}\,$GeV is fixed by the $SU(5)$ unification
of the gauge couplings, these relations in turn imply that
$V_H\sim 10^{17}\,$GeV and $M_P \sim 10^{18}\,$GeV, so that $M_P$ 
is indeed close to the string or Planck scale.
It is also noteworthy also that the scale $V_H\sim \sqrt{M_X M_P}$ 
could naturally emerge in the context of the models 
discussed in refs.\cite{PLB,BCL}.

\section{Conclusion } 
\vspace{-0.1cm}

In conclusion, I briefly discuss some actualities of the flavour 
problem:

$\bullet$ {\em How many families?} 
This difficult question is originated by the fact of family 
replication itself. There is no simple answer to the question 
``why three families?", or ``why {\em only} three families?".
In the SM (MSSM) as well as in GUTs the number of families $N_f$ 
remains to be an arbitrary parameter. Nevertheless, it seems that 
Nature prefers 3-family variant. 
The measurement of the $Z$-boson decay width indicates that 
there are no more standard-like families with light neutrinos. 
The standard model precision data constrain the number 
of families as $N_f<6$, so that only one or two extra heavy families 
with heavy ($M>M_Z/2$) neutrinos are still allowed.  

In the MSSM with very modest values of $\tan\beta$, 
there is still some (though not much) 
room for the fourth family\cite{Gunion}. 
Concerning SUSY GUTs, the presence of the fourth family 
would not affect the gauge coupling unification. 
However, this would affect the RG running of the Yukawa constants 
and thus spoil the {\em Grand Prix} of the $b-\tau$ unification. 
In the SUSY $SU(6)$ model there is no room for extra heavy family
since due to Goldstone nature of the Higgs doublets only one  
fermion, namely the top, can have $\sim 100$ GeV mass.

$\bullet$ {\em Neutrino Masses.} 
If the $SU(5)$ is a fundamental theory up to the Planck scale $M_P$, 
or if it is embedded in $SU(6)$ at some scale below $M_P$, 
then the natural value of the neutrino masses 
is  rather $\mcirc=v^2/M_P\sim 10^{-5}$ eV. This can solve 
the SNP via `just-so' oscillation\cite{just-so},  
but other neutrino hints\cite{ANP,LSND,HDM} are left unexplained. 
As for the $SO(10)$ theory, it contains the RH neutrinos $\nu^c$, 
and generates the LH neutrino masses by means of the seesaw 
mechanism\cite{seesaw}. 
The predictive $SO(10)$ frameworks\cite{ADHRS,BB,PLB} allow 
to calculate the neutrino Dirac masses, and their spectrum  
typically has the same shape as that of the charged fermions 
(see e.g. eqs. (\ref{ADHRS}) or (\ref{mf-1})). However, 
this cannot suffice for deducing the neutrino mass pattern, 
since now the question mark is contained in the RH neutrinos 
Majorana masses $M_R$.  The latter can emerge through the $d=5$ 
operators involving the the Higgs 16-plets $\psi,\bar{\psi}$: 
\be{RH} 
\frac{g_{ij}}{M_P} (16_i\bar{\psi})(\bar{\psi}16_j)
\ee 
Thus we obtain $M_R\sim M^2_{10}/M_P$, 
where $M_{10}$ is the $SO(10)\to SU(5)$ breaking scale 
which can range from $M_X\simeq 10^{16}$ GeV to $M_P=10^{19}$ GeV, 
and therefore $M_R\sim 10^{12-16}$ GeV. 
This corresponds to the mass of the heaviest neutrino ($\nu_\tau$) 
of about $10^{-3}-10$ eV, in which 
case the SNP can be due to the MSW mechanism\cite{MSW},  
and still some room can be left 
for explaining one of the other neutrino hints. 
For example, if $m_{\nu_\tau}\sim 0.1$ eV and 
$m_{\nu_\mu}\sim 10^{-2.5}$ eV, 
the SNP can be explained via the MSW oscillation 
$\nu_e\to nu_\mu$ while the ANP can be explained via 
the $\nu_\mu\to nu_\tau$ oscillation with large mixing angle. 
More precise predictions, however, would require to fix the pattern 
of constants $g_{ij}$ by imposing some symmetry constraints. 

It is worth to remark also that if all 
recent neutrino hints\cite{SNP,ANP,LSND,HDM}  
will be confirmed by future experiments, then three neutrino states 
$\nu_{e,\mu,\tau}$ will not suffice for thir explanation. 
It was shown in ref.\cite{calmoh} that only one possibility is 
compatible with all these data,  which requires an extra light sterile 
neutrino $\nu_s$ beyond the three known neutrinos. Then, assuming 
that $m_{\nu_{e,s}}\ll m_{\nu_{\mu,\tau}}$, the SNP can be explained 
by the $\nu_e\to nu_s$ oscillation and the ANP by the  
$\nu_\mu\to nu_\tau$ oscillation. In addition, almost degenerate 
$\nu_\mu$ and $\nu_tau$ with mass of about $2.5$ eV provide 
the cosmological HDM and can also explain the LSND oscillation 
$\bar{\nu}_\mu\to \bar{\nu}_e$. 

Of course, one needs to explain naturally the existence of so 
light sterile neutrino. According to the recent proposals, it 
could emerge in the particle spectrum as a light pseudo-Goldstone 
fermion\cite{CJS}, 
or as a neutrino of a hidden `mirror' world\cite{mirror}.    

$\bullet$ {\em Proton decay.} A realistic SUSY GUT must be 
capable to prevent proton from too fast decay. 
The gauge `dinosaur' boson mediated proton decays are not 
dangerous in SUSY GUT owing to the large unification scale 
$M_X\sim 10^{16}$ GeV. In SUSY GUTs proton wants to decay 
rather fastly via the $d=5$ operators\cite{dim5}:  
\be{LLLL} 
O_L=\frac{1}{M} q_iq_jq_kl_m, ~~~~~~ 
O_R=\frac{1}{M}u_i^c u_j^c d_k^c e_m^c  
\ee
which are mediated by exchanges of the heavy triplets $T,\bar T$ 
with mass $M\sim M_X$.  These couple to fermions as $q_iq_jT$, 
$q_kl_m\bar T$ etc., which couplings in SUSY GUT emerge  
on the same grounds as the ones of the Higgs doublets $\phi_{1,2}$
in eq. (\ref{Y-MSSM}). 
Although operators (\ref{LLLL}) involve the small Yukawa 
constants and mixing angles, they are anyway dangerous since they 
contain only one power of the large scale $M_X$. 
The operators $O_L$ dressed by Winos bring dominant 
contribution to the proton decay via the channel $p\to K\nu$.  
A dedicated analysis\cite{Nath} in the minimal $SU(5)$ model shows that for 
a typical parameter range proton lifetime in the above mode 
is about 2-3 orders of magnitude below the present experimental bound, 
and can be marginally reconciled with this bound if all inherent 
uncertainties are taken at extreme borders. 
In generic GUTs with nontrivial Clebsches the prediction of minimal 
$SU(5)$ are not valid. However, if there are no conspiracies, 
the problem still remains since the typical Clebsches about 2 or 3 
in general do not suffice to properly suppress proton decay. 

However, there are several theoretical possibilities for 
natural solution of this problem. 
The proton decay can be suppressed by the same mechanism 
which provides the predictivity for fermion masses. 
This happens, e.g. in the `inverse' hierarchy 
$SO(10)$ model\cite{SO10} where the dangerous operators 
$O_L$ vanish  automatically, while the more safe 
operators $O_R$ are left with a chance to show up in future 
experiments, essentially via the decay modes like  $p\to K\mu$, etc.   
There also exist more devoted models in which both operators 
$O_{L,R}$ can be killed by special arrangements in the Higgs 
or Fermion sectors\cite{BabuBarr,Valle,Dvali}.

Proton decay can be 
(at least partially) suppressed by the horizontal symmetry. 
For example, in the $SU(5)\times SU(3)_H$ model\cite{ZB2} 
the light up quarks ($u$ and $c$) get mass via antisymmetric 
operators $\frac{1}{M}10_\al \chi^{[\al\beta]} H 10_\beta$ 
($\al,\beta=1,2,3$ is a family index). 
On the other hand, the coupling $q_\al q_\beta T$ is symmetric, 
and thus it cannot emerge from antisymmetric operator. 
Then only $q_3 q_3 T$ is allowed, which emerges from the symmetric 
operator in (\ref{su3h-su5}), so that only the 
third family sfermions contribute the dominant decay mode 
$p\to K\nu$.  This can suppress the proton decay rate 
by about 2 orders of magnitude  
as compared to the minimal SUSY $SU(5)$ prediction.

$\bullet$ {\em Sparticle spectrum and flavour changing phenomena.} 
In the MSSM the universal soft SUSY breaking\cite{BFS} 
guarantees natural suppression of the FCNC phenomena 
mediated by the sparticles. However, in the context of 
SUSY GUT this becomes insufficient, since the impact of 
physics betweem the Planck and GUT scales can strongly violate 
the soft-terms universality\cite{HKR,BH}.  

Predictive SUSY GUT frameworks, generically based on the 
HFE mechanism, employ a rich fermion sector ($F$-fermions) above 
the GUT scale, and  all these fermions typically have large $O(1)$   
Yukawa couplings. Then the  soft masses of different sfermions 
with the same standard charges will have different RG running down 
from the scale $M_P$, so that at the scale $M_X$ where the heavy  
states decouple from the light ones, the masses of the latter 
are no more universal\cite{DP}. 
As a result, at lower energies the sfermions can arrive being already 
substantially split between different families, which would give 
rise to the FCNC phenomena. 

Let us consider, as an example, the case of the $SU(6)$ model 
discussed in previous section.  The FCNC problem emerges e.g. 
due to the third term in eq. (\ref{Yukawa}), even if it 
does not contribute the light fermion masses and 
in fact works only for rendering the extra $(5+\bar 5')_i$ states 
superheavy. 
The Yukawa couplings $\Ga_{ij}\sim 1$, unless they are degenerate, 
would cause different RG running of particle masses 
from $M_P$ down to the $SU(6)$ symmetry breaking scale $V_H$. 
Then  the soft masses e.g. of sleptons $\tilde{e}_i^c$, contained 
in the multiplets $15_i$ ($i=1,2,3$), should be strongly split 
already at the scale $V_H$. The 
non-universality of the slepton masses would induce decays 
like $\mu\to e+\ga$, EDM of the electron, etc.\cite{BH}. 

The flavour-changing problem is a challenge for the relations 
of the two (\ref{Love}). In particular, splitting 
between the sfermion masses of the first two families 
would cause very serious problems, since in this case the 
$\mu\to e+\ga$ decay rate, the rates of the $K^0-\bar K^0$ 
transition, etc. would strongly exceed the experimental bound.  

Besides the disorientation and plastification introduced in 
particle physics by Dimopoulos et al.\cite{Gian}, an idea of 
the non-Abelian horizontal symmetry, say $SU(3)_H$,
can be a natural way for solving the problem.  
For example, the model of ref.\cite{ZB2} (see section section 2) 
can naturally render the sfermion masses unsplit (at least 
within first two families) in spite the renormalization effects 
caused by the large Yukawa constants.  Since in this model 
the horizontal $SU(2)_H$ subgroup between the first two families 
is broken below the scale $M_X$, soft masses of the latter 
will have the same RG evolution down to the scale $M_X$ 
and thus will remain unsplit.  
They will not get strong splitting due to mixing effects with the 
heavy $F$-states $\bar{\rm X} + {\rm X}$ and ${\rm V} + \bar{\rm V}$ 
neither, since the mixing angles are small: 
of the order of corresponding Yukawa constants 
in MSSM, $\la_c,\la_s$, etc.   
As for the third family fermions, their soft masses in general should 
be split from the ones of first two generations,  
since $\la_t\sim 1$ and thus they are strongly mixed to the soft 
masses of the corresponding $F$-states.   
However, splitting between the third and the first two 
families of sfermions is not yet a problem\cite{BH}. 

\vspace{10mm}

{\bf References} 

\newpage

\baselineskip=12pt

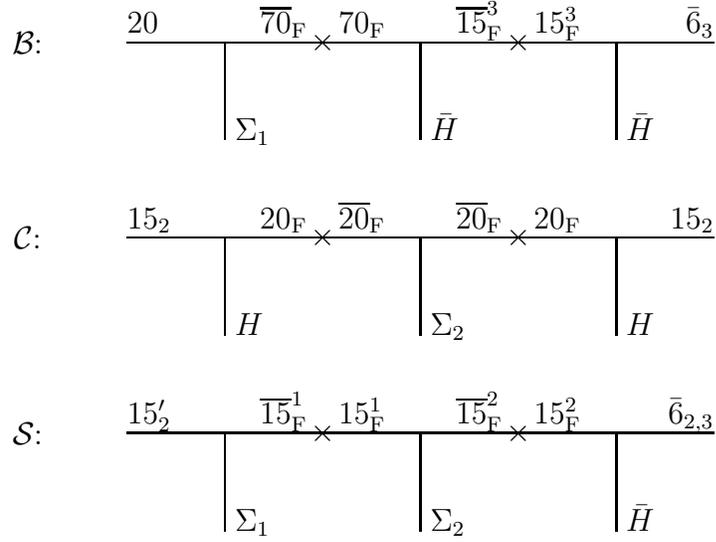
\begin{figure}\setlength{\unitlength}{1.3cm}
\begin{center}\begin{picture}(6,5.7)(0,-6)
\put(-1,-0){\makebox(0,0){ ${\cal B}$: }}
\put(-1,-2){\makebox(0,0){ ${\cal C}$: }}
\put(-1,-4){\makebox(0,0){ ${\cal S}$: }}
\put(0,0){\line(1,0){6}}
\multiput(1,0)(2,0){3}{\line(0,-1){1}}
\put(0,0.1){\makebox(0,0)[bl]{20}}
\put(2,0.1){\makebox(0,0)[b]{$\overline{70}_{\rm F}~~~70_{\rm F}$}}
\put(4,0.1){\makebox(0,0)[b]{$\overline{15}^3_{\rm F}~~~15^3_{\rm F}$}}
\put(6,0.1){\makebox(0,0)[br]{$\bar{6}_3$}}
\put(1.1,-1){\makebox(0,0)[bl]{$\Sigma_1$}}
\put(3.1,-1){\makebox(0,0)[bl]{$\bar{H}$}}
\put(5.1,-1){\makebox(0,0)[bl]{$\bar{H}$}}
\put(2,0){\makebox(0,0){$\times$}}
\put(4,0){\makebox(0,0){$\times$}}
\put(0,-2){\line(1,0){6}}
\multiput(1,-2)(2,0){3}{\line(0,-1){1}}
\put(0,-1.9){\makebox(0,0)[bl]{$15_2$}}
\put(2,-1.9){\makebox(0,0)[b]{$20_{\rm F}~~~\overline{20}_{\rm F}$}}
\put(4,-1.9){\makebox(0,0)[b]{$\overline{20}_{\rm F}~~~20_{\rm F}$}}
\put(6,-1.9){\makebox(0,0)[br]{$15_2$} }
\put(1.1,-3){\makebox(0,0)[bl]{$H$} }
\put(3.1,-3){\makebox(0,0)[bl]{$\Sigma_2$}}
\put(5.1,-3){\makebox(0,0)[bl]{$H$}}
\put(2,-2){\makebox(0,0){$\times$}}
\put(4,-2){\makebox(0,0){$\times$}}
\put(0,-4){\line(1,0){6}}
\multiput(1,-4)(2,0){3}{\line(0,-1){1}}
\put(0,-3.9){\makebox(0,0)[bl]{$15'_2$} }
\put(2,-3.9){\makebox(0,0)[b]{$\overline{15}^1_{\rm F}~~~15^1_{\rm F}$}}
\put(4,-3.9){\makebox(0,0)[b]{$\overline{15}^2_{\rm F}~~~15^2_{\rm F}$}}
\put(6,-3.9){\makebox(0,0)[br]{$\bar{6}_{2,3}$}}
\put(1.1,-5){\makebox(0,0)[bl]{$\Sigma_1$}}
\put(3.1,-5){\makebox(0,0)[bl]{$\Sigma_2$}}
\put(5.1,-5){\makebox(0,0)[bl]{$\bar{H}$}}
\put(2,-4){\makebox(0,0){$\times$}}
\put(4,-4){\makebox(0,0){$\times$}}
\end{picture}
\caption{diagrams giving rise\label{Diagrams}
to the operators ${\cal B,~ C,~ S}$ }
\end{center}\end{figure}


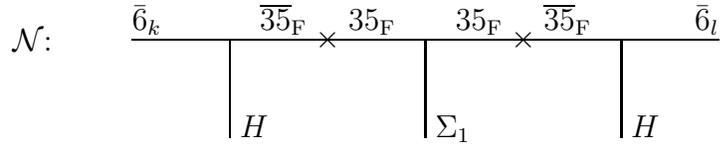
\begin{figure}\setlength{\unitlength}{1.3cm}
\begin{center}\begin{picture}(6,2)(0,-1)
\put(-1,-0){\makebox(0,0){ ${\cal N}$: }}
\put(0,0){\line(1,0){6}}
\multiput(1,0)(2,0){3}{\line(0,-1){1}}
\put(0,0.1){\makebox(0,0)[bl]{$\bar{6}_k$}}
\put(2,0.1){\makebox(0,0)[b]{$\ov{35}_{\rm F}~~~~35_{\rm F}$}}
\put(4,0.1){\makebox(0,0)[b]{$35_{\rm F}~~~~\ov{35}_{\rm F}$}}
\put(6,0.1){\makebox(0,0)[br]{$\bar{6}_l$}}
\put(1.1,-1){\makebox(0,0)[bl]{$H$}}
\put(3.1,-1){\makebox(0,0)[bl]{$\Sigma_1$}}
\put(5.1,-1){\makebox(0,0)[bl]{$H$}}
\put(2,0){\makebox(0,0){$\times$}}
\put(4,0){\makebox(0,0){$\times$}}
\end{picture}\caption{the diagram
giving rise to the operator ${\cal N}$ for neutrino mass \label{FalseMass}}
\end{center}\end{figure}


\begin{figure}\setlength{\unitlength}{1.3cm}
\begin{center}\begin{picture}(6,4)(0,-3.3)
\put(-1,-0){\makebox(0,0){ ${\cal D}$: }}
\put(-1,-2){\makebox(0,0){ ${\cal U}$: }}
%
\put(0,0){\line(1,0){8}}
\multiput(1,0)(2,0){4}{\line(0,-1){1}}
\put(0,0.1){\makebox(0,0)[bl]{$15_{i}$}}
\put(2,0.1){\makebox(0,0)[b]{$\overline{105}_{\rm F}~~105_{\rm F}$}}
\put(4,0.1){\makebox(0,0)[b]{$\overline{210}_{\rm F}~~210_{\rm F}$}}
\put(6,0.1){\makebox(0,0)[b]{$\overline{84}_{\rm F}~~~84_{\rm F}$}}
\put(8,0.1){\makebox(0,0)[br]{$\bar{6}_{k}$}}
\put(1.1,-1){\makebox(0,0)[bl]{$\Sigma_1$} }
\put(3.1,-1){\makebox(0,0)[bl]{$\bar{H}$} }
\put(5.1,-1){\makebox(0,0)[bl]{$\Sigma_1$} }
\put(7.1,-1){\makebox(0,0)[bl]{$\Sigma_1$} }
\put(2,0){\makebox(0,0){$\times$}}
\put(4,0){\makebox(0,0){$\times$}}
\put(6,0){\makebox(0,0){$\times$}}
\put(0,-2){\line(1,0){8}}
\multiput(1,-2)(2,0){4}{\line(0,-1){1}}
\put(0,-1.9){\makebox(0,0)[bl]{$15_{i}$}}
\put(2,-1.9){\makebox(0,0)[b]{$\overline{105}_{\rm F}~~105_{\rm F}$}}
\put(4,-1.9){\makebox(0,0)[b]{$20^1_{\rm F}~~~20^2_{\rm F}$}}
\put(6,-1.9){\makebox(0,0)[b]{$\overline{20}_{\rm F}~~~20_{\rm F}$}}
\put(8,-1.9){\makebox(0,0)[br]{$15_2$}}
\put(1.1,-3){\makebox(0,0)[bl]{$\Sigma_1$} }
\put(3.1,-3){\makebox(0,0)[bl]{$H$} }
\put(5.1,-3){\makebox(0,0)[bl]{$\Sigma_1$} }
\put(7.1,-3){\makebox(0,0)[bl]{$H$} }
\put(2,-2){\makebox(0,0){$\times$}}
\put(4,-2){\makebox(0,0){$\times$}}
\put(6,-2){\makebox(0,0){$\times$}}
\end{picture}
\caption{diagrams giving rise\label{Diagrams2}
to the operators ${\cal D}$ and ${\cal U}$ }
\end{center}\end{figure}
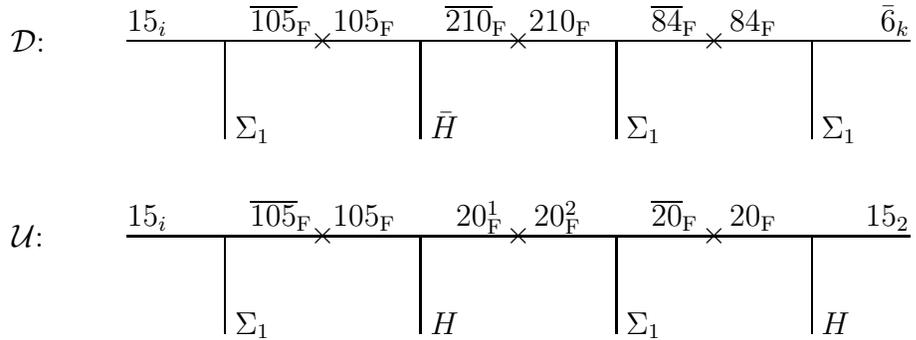


\end{document}